\documentclass[fleqn,usenatbib]{mnras}
\usepackage{newtxtext,newtxmath}
\usepackage[T1]{fontenc}
\usepackage{ae,aecompl}
\usepackage{graphicx}
\usepackage{dcolumn}
\usepackage{bm}
\usepackage{color}
\usepackage{natbib}
\usepackage{ulem}
\usepackage{multirow}

\usepackage{xspace}
\usepackage{nicefrac}
\newcommand{\KEPLER}{{\textsc{Kepler}}\xspace}

\newcommand{\A}[1]{\ensuremath{A(\text{#1})}\xspace}
\newcommand{\Abold}[1]{\ensuremath{\boldmath{A(\text{#1})}}\xspace}
\newcommand{\B}[2]{\ensuremath{[\text{#1}/\text{#2}]}\xspace}
\newcommand{\Msun}{\ensuremath{\mathrm{M}_\odot}}
\newcommand{\Lsun}{\ensuremath{\mathrm{L}_\odot}}
\newcommand{\Ep}[1]{\ensuremath{10^{#1}}}
\newcommand{\E}[1]{\ensuremath{\times\Ep{#1}}}

\newcommand{\cm}{\ensuremath{\mathrm{cm}}}

\newcommand{\erg}{\ensuremath{\mathrm{erg}}}
\newcommand{\g}{\ensuremath{\mathrm{g}}}
\newcommand{\K}{\ensuremath{\mathrm{K}}}
\newcommand{\yr}{\ensuremath{\mathrm{yr}}}
\newcommand{\kyr}{\ensuremath{\mathrm{kyr}}}
\newcommand{\vcrit}{\ensuremath{v_\mathrm{crit}}}
\newcommand{\vrot}{\ensuremath{v_\mathrm{rot}}}
\newcommand{\fmu}{\ensuremath{f_\mu}}
\newcommand{\fc}{\ensuremath{f_\mathrm{c}}}

\title{Rapidly Rotating Massive Pop III stars: A Solution for High Carbon Enrichment in CEMP-no Stars}

\author[S. K. Jeena et al.]{
S. K. Jeena,$^{1}$\thanks{E-mail: jeenaunni44@gmail.com}
Projjwal Banerjee,$^{1}$ 
Gen Chiaki,$^{2,3}$
and Alexander Heger$^{4,5,6}$
\\
$^{1}$Department of Physics,  Indian Institute of Technology Palakkad, Kerala, India\\
$^{2}$Astronomical Institute, Graduate School of Science, Tohoku University, Aoba, Sendai 980-8578, Japan\\
$^{3}$National Astronomical Observatory of Japan, 2-21-1 Osawa, Mitaka, Tokyo 181-8588, Japan \\
$^{4}$School of Physics and Astronomy, Monash University, Vic 3800, Australia\\
$^{5}$OzGrav: The ARC Centre of Excellence for Gravitational Wave Discovery, Australia\\
$^{6}$ARC Centre of Excellence for Astrophysics in Three Dimensions (ASTRO-3D), Australia\\
}

\date{Accepted XXX. Received YYY; in original form ZZZ}

\pubyear{2018}
\begin{document}
\label{firstpage}
\pagerange{\pageref{firstpage}--\pageref{lastpage}}
\maketitle
\begin{abstract}
Very metal-poor stars that have $[\text{Fe}/\text{H}]<-2$ and that are enhanced in C relative to Fe ($[\text{C}/\text{Fe}]>+0.7$) but have no enhancement of heavy elements ($[\text{Ba}/\text{Fe}]<0$) are known as carbon-enhanced metal-poor (CEMP-no) stars.  These stars are thought to be produced from a gas that was polluted by the supernova (SN) ejecta of the very first generation (Pop III) massive  stars. The very high enrichment of C ($A(\text{C})\gtrsim 6$) observed in many of the CEMP-no stars  is difficult to explain by current models of SN explosions from massive Pop III stars when a reasonable dilution of the SN ejecta, that is consistent with detailed simulation of metal mixing in minihaloes, is adopted.  We explore rapidly rotating Pop III stars that undergo efficient mixing and reach a quasi-chemically homogeneous (QCH) state.  We find that QCH stars can eject large amounts of C in the wind and that the resulting dilution of the wind ejecta in the interstellar medium can lead to a C enrichment of $A(\text{C})\lesssim7.75$.  
The core of QCH stars can produce up to an order of magnitude of more C than non-rotating progenitors of similar mass and the resulting SN can lead to a C enrichment of $A(\text{C})\lesssim7$.   
Our rapidly rotating massive Pop III stars cover almost the entire range of $A(\text{C})$ observed in CEMP-no stars and are a promising site for explaining the high C enhancement in the early Galaxy.  Our work indicates that a substantial fraction of Pop III stars were likely rapid rotators.
\end{abstract}


\section{Introduction}

Very metal-poor (VMP) stars ($\B{Fe}{H}< -2$) of mass $\lesssim0.8\,\Msun$ are thought to be the fossil records of the nucleosynthesis in the earliest generation of massive stars that were present in the early Galaxy.  A large fraction ($\sim 20\,\%$) of VMP stars are found to be enhanced in C relative to Fe ($\B{C}{Fe}>+0.7$) and are referred to as carbon-enhanced metal-poor (CEMP) stars \citep{beers2005,aoki2007carbon} with the frequency increasing rapidly with decreasing metallicity below $\B{Fe}{H}< -2$ \citep{Lucatello2006ApJ,lee2013,yong2013,Placco2014Apj}.  CEMP stars are found throughout the Milky Way halo \citep{carollo2012} as well as in dwarf spheroidal  and ultra-faint dwarf galaxies \citep{Norris2010ApJ,lai2011,skuladottir2015A&A,susmitha2017,chiti2018}. 
CEMP stars are further classified into CEMP-\textsl{s}, CEMP-\textsl{r/s}, CEMP-\textsl{r}, and CEMP-no stars based on the enrichment (\B{Ba}{Fe}) and abundance pattern (\B{Ba}{Eu}) of heavy elements \citep{beers2005}.  We list the details of the classification of CEMP stars in Table~\ref{tab:2}. 
The majority of CEMP stars are found to be either CEMP-\textsl{s,r/s} or CEMP-no stars. 
Although the origins of CEMP-\textsl{s} and CEMP-\textsl{r/s} stars are still being investigated, the most popular models to explain the high C and heavy element enrichment involve mass transfer from a low or intermediate mass binary companion.  Thus, it is believed that the observed surface abundance pattern of CEMP-\textsl{s} and CEMP-\textsl{r/s} do not reflect the composition of the interstellar medium (ISM) from which the stars were born. Whereas slow neutron capture (\textsl{s}-process) in low-mass stars during the asymptotic giant branch (AGB) phase can match the abundance pattern in CEMP-\textsl{s} stars \citep{bisterzo2}, intermediate neutron capture (\textsl{i}-process) is required to explain most of the CEMP-\textsl{r/s} stars \citep{hampel2016}.  The site for \textsl{i}-process is still under debate although a number of sites associated with the end stages of low and intermediate stars have been proposed \citep{fujimoto2000,campbell2010,herwig2011,deni2017}.
It is possible, however, that some of the CEMP-\textsl{s} and CEMP-\textsl{r/s} stars formed directly from the ISM polluted by early massive stars that underwent \textsl{s} and \textsl{i}-process \citep{frischk2016,choplin2017,banerjee2018,banerjee2018cemp,banerjee2019}.  In contrast, CEMP-no stars, that have a low abundance of heavy elements ($\B{Ba}{Fe}<0.0$), are thought to be produced from the ISM polluted by the supernova (SN) ejecta of the earliest generation of massive stars including the first generation (Pop III) stars \citep{spite2013A&A,norris2013,Placco2014Apj}.
\begin{table}
    \centering
    \caption[Classification of CEMP stars]{Classification of CEMP stars according to~\citet{beers2005}.}
    \begin{tabular}{l l}
    \hline
     
    CEMP-\textsl{r}     &[C/Fe]$>$+0.7 and [Ba/Eu]$<$0.0\\
   CEMP-\textsl{s} & [C/Fe]$>$+0.7, [Ba/Fe]$>$+1.0,\\ &  and [Ba/Eu]$>$+0.5\\
    CEMP-\textsl{r/s} & [C/Fe]$>$+0.7 and 0.0$<$[Ba/Eu]$<$+0.5\\
    CEMP-no & [C/Fe]$>$+0.7 and [Ba/Fe]$<$0.0\\
    \hline
    
    \end{tabular}
    
    \label{tab:2}
\end{table}

\begin{figure}
     \centering
     \includegraphics[ width=\columnwidth]{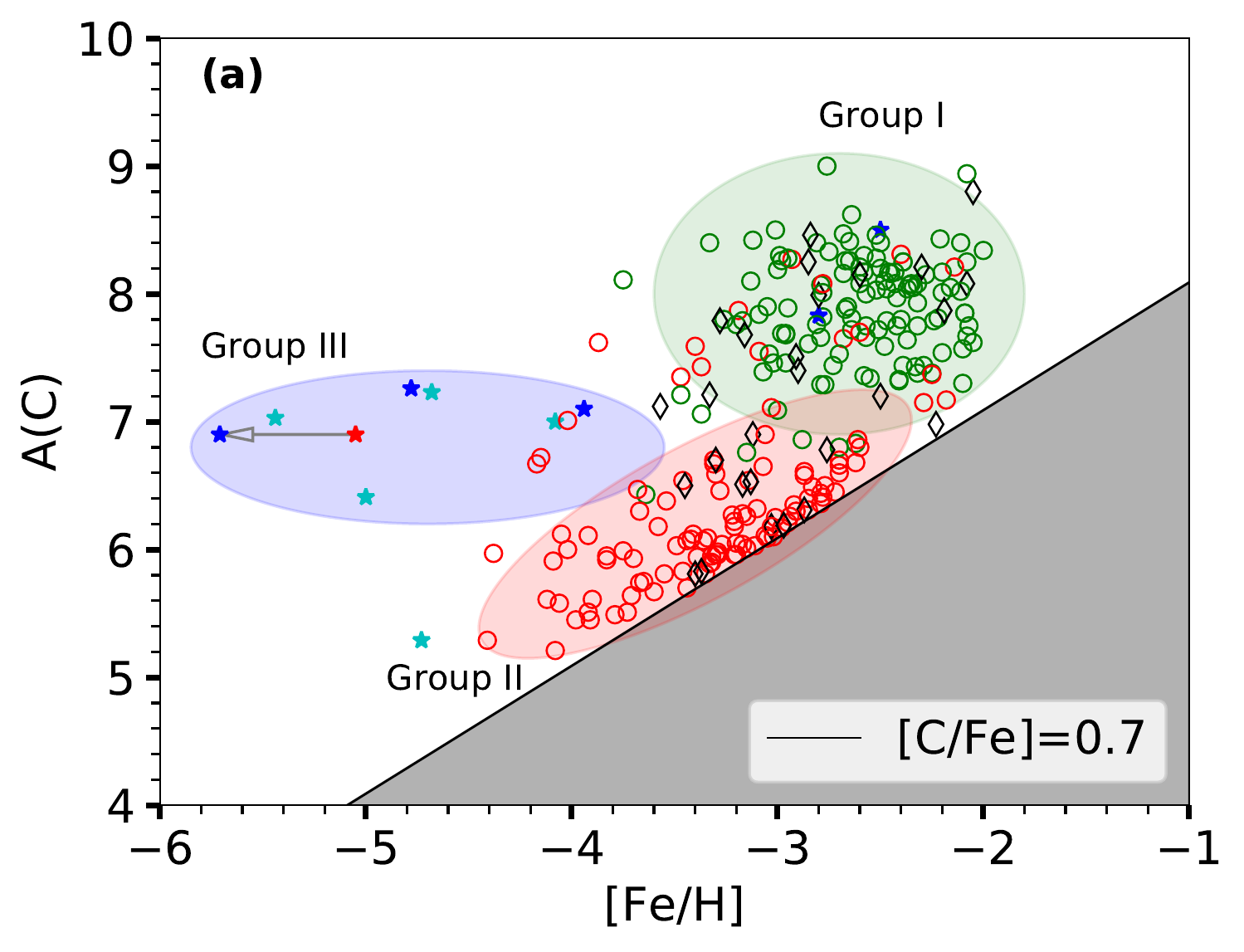} 
     \includegraphics[ width=\columnwidth]{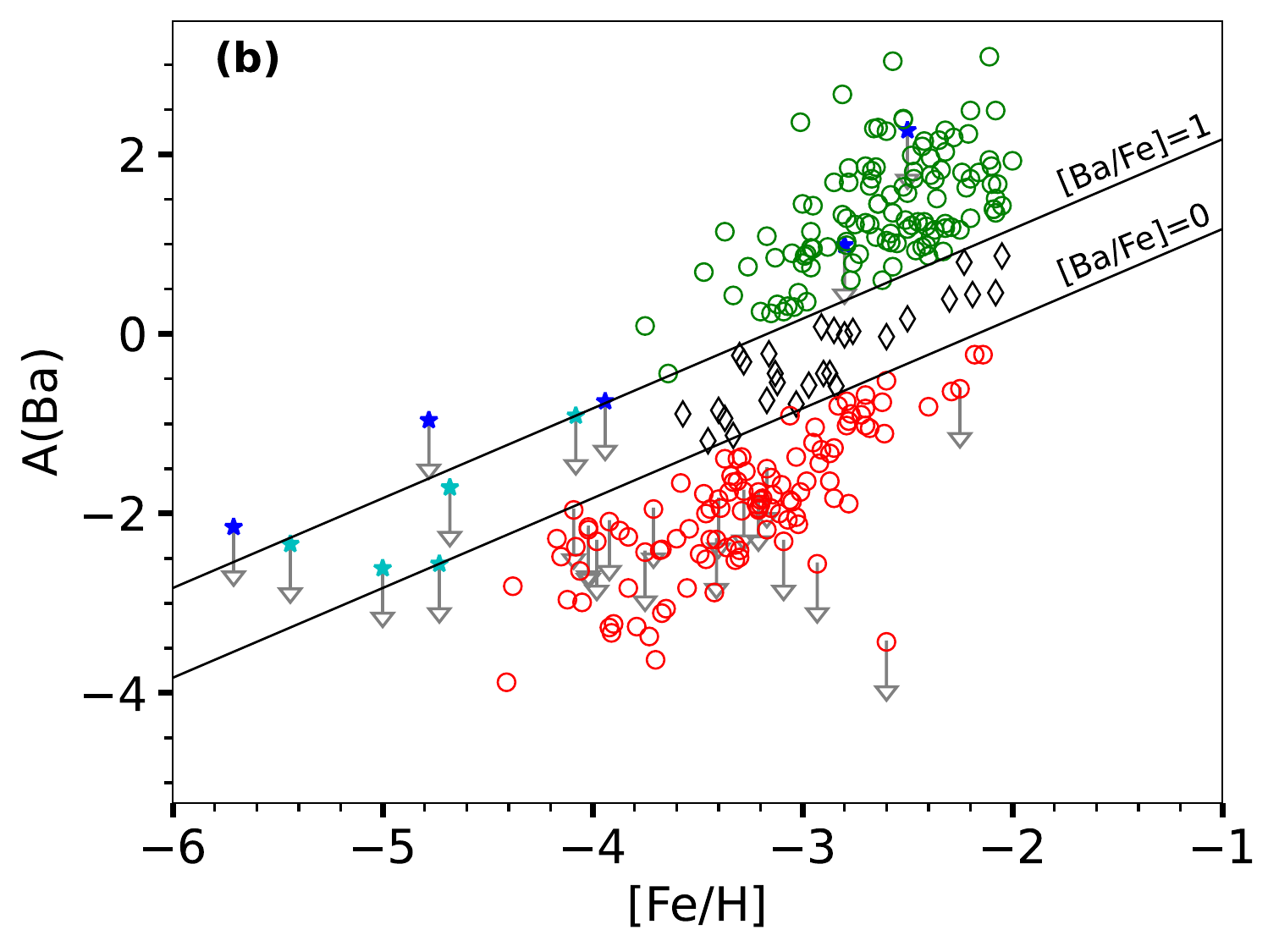} 

    \caption[Yoon-Beers diagram: classification of CEMP stars]{(a) Yoon-Beers diagram: classification of CEMP stars according to ~\citet{yoon2016observational} based on the \A{C} Vs \B{Fe}{H} plot.  The \textsl{red} and \textsl{green open circles} are for CEMP-no and CEMP-\textit{s} stars, respectively.  The \textsl{black diamonds} indicate stars with $0<\B{Ba}{Fe}<1$.  The \textsl{cyan-filled stars} show CEMP stars in the lower metallicity regime ($\B{Fe}{H}<-4$) with an upper limit of Ba measurement and with $0\leq\B{Ba}{Fe}_\mathrm{upper}\leq1$.  The \textsl{blue asterisks} represent stars with $\B{Ba}{Fe}_\mathrm{upper}>1$.  The \textsl{solid red asterisks} represents stars with an upper limit of \B{Fe}{H} but without Ba measurement.  The \textsl{black inclined line} provides a reference at $\B{C}{Fe}=0.7$. (b) The corresponding plot of \A{Ba} Vs \B{Fe}{H} for the CEMP stars are plotted in (a). The \textsl{black inclined lines} provide references at $\B{Ba}{Fe}=0.0$ and $1$.} 
     \label{fig:yoon1}
 \end{figure}

An interesting feature in CEMP-no stars was found when the absolute abundance \A{C}\footnote{$\A{X}=\log\epsilon(X)\equiv\log(N_X/N_\mathrm{H})+12$, where $N_X$ and $N_\mathrm{H}$ are number abundance of element $X$ and H, respectively.} is plotted versus \B{Fe}{H} \citep{Bonifacio2015A&A,yoon2016observational}. The resulting plot is referred to as the Yoon-Beers (YB) diagram. Figure~\ref{fig:yoon1}a shows the YB diagram for the sample of CEMP stars considered in~\citet{yoon2016observational} where open red circles are for CEMP-no stars and open green circles are for CEMP-\textsl{s} and CEMP-\textsl{r/s} stars which we will refer to as just CEMP-\textsl{s} stars for simplicity. We also plot stars with $0<\B{Ba}{Fe}<1$ as open black diamonds which cannot strictly be classified as either CEMP-no or CEMP-\textsl{s}. Additionally, we use filled asterisks for stars with an upper limit of Ba where cyan and blue filled asterisks represent $0\leq\B{Ba}{Fe}_\mathrm{upper}\leq1$ and $\B{Ba}{Fe}_\mathrm{upper}>1$, respectively. These stars also cannot be classified as either CEMP-no or CEMP-\textsl{s} stars.  As can be seen from Fig.~\ref{fig:yoon1}a, the majority of the CEMP-\textsl{s} stars have a high C abundance of $7\lesssim\A{C}\lesssim9$ and $\B{Fe}{H}\gtrsim-3.5$ and are referred to as Group I stars.  On the other hand, for the majority of CEMP-no stars, \B{Fe}{H} and \A{C} are roughly correlated for values ranging from $-5\lesssim\B{Fe}{H}\lesssim-2.5$ with $5\lesssim\A{C}\lesssim 7$.  These are referred to as Group II stars. Lastly, at very low metallicity of [Fe/H]$\lesssim -3.5$ there are some stars that have high C enrichment of $6.2 \lesssim \A{C} \lesssim 7.3$ that is not correlated with \B{Fe}{H}, and are referred to as Group III stars.  
Most of the stars in Group III have an upper limit of Ba where  $\B{Ba}{Fe}_\mathrm{upper}>0$ (filled asterisks) and thus cannot be classified as CEMP-no or CEMP-\textsl{s} stars.

Figure~\ref{fig:yoon1}b shows \A{Ba} as a function of \B{Fe}{H}. As can be seen from the figure, for $\B{Fe}{H}\gtrsim-4$, CEMP-\textsl{s} and CEMP-no stars are clearly separated from each other where the former has $\A{Ba}\gtrsim0$ and the latter has $\A{Ba}<0$.  Stars that have values of $0\leq\B{Ba}{Fe}<1$ (black diamonds) lie in between the two clusters separating CEMP-\textsl{s} and CEMP-no stars.  With regard to Group III, although most of the stars only have an upper limit of observed Ba and cannot strictly be classified as CEMP-no stars, Ba enrichment is limited to very low values of $\A{Ba}\lesssim-0.5$ which is distinct from the values found in CEMP-\textsl{s} stars that have $\A{Ba}\gtrsim 0$.  For this reason, similar to \citet{yoon2016observational}, we will also refer to such stars in Group III as CEMP-no stars along with normal CEMP-no stars that have $\B{Ba}{Fe}<0$ (red circles in Fig.~\ref{fig:yoon1}).

The extremely low metallicity of Group III stars indicates that they are very likely associated with the nucleosynthesis products of massive Pop III stars. With regard to Groups II and I, a recent chemodynamical study by \citet{zepeda2023} of a large number of CEMP stars indicates that Group II stars are likely born from the ISM polluted by Pop III and early massive stars whereas the C enhancement in Group I stars is a result of local phenomenon such as mass transfer from a binary companion.

\subsection{Origin of CEMP-no stars}
Assuming that CEMP-no stars are formed from the ISM polluted by only one --- or at least very few --- early-generation massive stars, we can directly use the surface abundance pattern of the CEMP-no stars to constrain the properties of the Pop III stars and their SNe. 
Among the several models that have been proposed to explain the abundance pattern observed in these stars, models that are associated with faint SNe are the most popular \citep{umeda2003first,umeda2005variations,heger2010nucleosynthesis,tominaga2014abundance}.  These models use the ``mixing and fallback'' mechanism in which a large fraction of the inner ejecta falls back onto the remnant neutron star or black hole. As a result, the ejection of light elements, such as C, that are present in the outer layers, dominates over the heavier (Fe group) elements that are present in the innermost layers.  Such ejecta have average values of $\B{C}{Fe}\gtrsim 0.7$.

Another model that has been proposed to explain CEMP-no stars are associated with rapidly rotating massive stars, known as ``spinstars" \citep{meynet2006early,meynet2010c,chiappini2013first}.  In this scenario, rotation-induced mixing leads to the enhancement of the envelope with CNO isotopes that are subsequently ejected during the explosion.

\subsection{Minimum Dilution and \Abold{C} Problem}
The existing models for CEMP-no stars mentioned above can explain the relative abundance patterns in many of the CEMP-no stars, including the high values of C relative to Fe.  In order to match the corresponding absolute C abundances, however, the level of dilution of the SN ejecta adopted in these models is incompatible with the dilution found in simulations of SN metal mixing from Pop III stars.   In a homogeneous mixing scenario, the SN ejecta is diluted with a minimum amount of mass in the ISM $M_{\rm dil}^{\min}$ corresponding to the swept-up mass of a spherically symmetric blast wave until the shock speed becomes comparable to the sound speed.  In reality, this is just the lower bound of the level of dilution as further dilution due to turbulent mixing and gas inflow as well as inhomogeneous mixing of SN ejecta with the ISM will result in larger effective dilution before the next generation of stars is formed.  This is particularly relevant for SN resulting from Pop III stars as they are formed in small dark matter minihaloes of $\sim2$--$30\E5\,\Msun$ \citep{abel1998first,abel2002formation}.  It was pointed out recently by \citet{magg2020minimum} that in a large number of studies that simulate detailed inhomogeneous metal mixing of SN ejecta and subsequent collapse of gas into star-forming regions in minihaloes, the effective levels of dilution are always larger than a minimum dilution mass $\sim2\E4\,\Msun$ and in most cases orders of magnitude higher. The particular value of 
\begin{equation}
M_{\rm dil}^{\rm min} \simeq 
2.4\E4\,\Msun
\left( \frac{E}{\Ep{51}\,\erg}\right) ^{0.96}
\left( \frac{n_0}{0.1\,\cm^{-3}} \right)^{-0.11},
\label{eq:magg}
\end{equation}
corresponds to the mass swept-up of a spherically symmetric SN blast wave of energy $E \sim\Ep{51}\,\erg$ in a fully ionized ISM of ambient density $n_0 \sim 0.1\,\cm^{-3}$ before it fades away \citep{magg2020minimum}. 

A detailed study of metal enrichment in minihaloes resulting from core-collapse SNe (CCSNe) and pair-instability SNe (PISNe) resulting from exploding Pop III stars by \citet{chiaki2018} found that the enrichment level of next-generation star-forming regions depends on several factors such as the amount of photoionizing radiation during the life of the star, the dark matter mass of the minihalo, and the progenitor mass.  Overall, metal enrichment in minihaloes can be broadly divided into internal enrichment and external enrichment.  In the case of internal enrichment, SN ejecta is confined to the host minihalo and recollapses back into star-forming regions of the host minihalo, whereas, for external enrichment, the ejecta escapes the host minihalo and pollutes a neighbouring halo.  External enrichment, which occurs mainly in the PISN cases, is found to be highly inefficient with an effective dilution of $\gtrsim\Ep9\,\Msun$. 
On the other hand, for internal enrichment, which occurs mainly in the CCSN cases, although the efficiency of enrichment is higher than external enrichment, the effective dilution mass covers a wide range from $\sim3.5\E4$--$8\E7\,\Msun$ with average values of $\gtrsim\Ep5\,\Msun$ for a CCSN of energy of $\Ep{51}\,\erg$ resulting from progenitors of $13-30\,\Msun$.  A similar level of dilution of $\sim 2.5\E5\,\Msun$ was again found by \citet{chiaki2019seeding} resulting from a CCSN of energy $\Ep{51}\,\erg$ from a $13.5\,\Msun$ Pop III progenitor.  In another study by \citet{chiaki2020seeding} that specifically looked at the formation of CEMP stars from faint SNe from Pop III progenitors of mass $13$--$80\,\Msun$, it was found that the C enrichment ranges from $\A{C}\sim4$--$5$ with very low $\B{Fe}{H}\lesssim-8$ and consequently cannot explain any of the observed CEMP-no stars.  Thus, even in the most optimistic scenario corresponding to an effective minimum dilution of $\sim 3.5\E4\,\Msun$, regular as well as faint CCSN resulting from non-rotating Pop III progenitors of $\lesssim30\,\Msun$, that have C yields of $\lesssim0.4\,\Msun$, can at most lead to $\A{C}\le6.1$ with typical values of $\lesssim5$.
Similar values were also obtained from a semi-analytical model of chemical evolution by \citet{komiya2020ApJ} where the maximum value of $\A{C}$ for $\B{Fe}{H}<-3$ was found to be $\sim 5.6$ ($\B{C}{H}\sim-2.8$) when fiducial values of $M_{\rm dil}^{\rm min}=5.1\E4\,\Msun$ and turbulent mixing parameter were adopted.  In a recent study of a cosmological hydrodynamic zoom-in simulation of an isolated ultra-faint dwarf, \citet{jeon2021MNRAS.} also find that it is challenging to explain CEMP-no stars with $\A{C}\gtrsim7$ even with faint SNe.  They, however, report somewhat larger values of $\A{C}\gtrsim 6$ in their simulation which is likely due to the use of high mass non-rotating Pop III stars of $50$--$100\,\Msun$ that can eject higher amounts of C of up to $\sim 1.5\,\Msun$ but are unlikely to explode~\citep{muller2016simple}. 

In all the simulations of metal enrichment in minihaloes, only non-rotating Pop III progenitors were considered.  As mentioned earlier, however,  rapidly rotating ``spinstar'' models have also been proposed as a potential source explaining CEMP-no stars \citep{meynet2006early,meynet2010c,chiappini2013first,choplin2017pre}.
In these models, however, although the wind is enriched by CNO-processed material and substantial amounts of $^{13}$C and $^{14}$N are ejected in the wind, 
very little $^{12}$C is ejected ($\lesssim0.02\,\Msun$). 
In order to eject a substantial amount of $^{12}$C, most of the core has to be ejected during the SN explosion.  This would again lead to similar levels of dilution as for regular CCSN from non-rotating progenitors leading to similar values of \A{C}.  In a recent study \citet{liu2021} have explored the ``spinstar'' models with \textit{ad hoc} mass loss prescription via wind that result in tremendous mass loss.  The mass loss ranges from a  minimum value of all material above the He core to a maximum value of all material up to the central remnant, which essentially covers the entire core of the star. \citet{liu2021} find that the resulting enrichment when substantial material from the O/C core is ejected can lead to C enrichment of $\A{C}\sim 7$.  Such dramatic mass loss, however, that ejects almost the entire mass of the star up to the central remnant is very unlikely.

Thus, all of the existing models of CEMP-no stars only cover a small fraction of space in Group II and Group III stars in the YB diagram (Fig.~\ref{fig:yoon1}a). The rest of the CEMP-no stars in Group II and Group III are difficult to explain with the existing scenarios.
We explore rapidly rotating models of Pop III stars that undergo efficient mixing and reach the so-called quasi-chemically homogeneous (QCH) state.  We find that rapidly rotating models that reach the QCH state can eject substantial amounts of CNO in the wind as well as during their explosion which can explain the absolute C enrichment observed in CEMP-no stars even when a reasonable value of minimum dilution is adopted. This makes them a promising site for explaining the origin of CEMP-no stars.

The layout of the paper is as follows: In Section~\ref{sec:method}, we briefly describe the methods used for the models.  The details of the evolution of rapidly rotating massive star models that reach the QCH state and associated results are discussed in Section~\ref{sec:evolution and results}.  In Section~\ref{sec:discussion}, we compare the ejecta from our models with the detailed observed abundance patterns of CEMP-no stars.  Finally, we conclude with a summary of the paper in Section~\ref{sec:summary}.

\section{Method}\label{sec:method}

We explore models of rapidly rotating massive stars using the 1D hydrodynamic stellar evolution code \KEPLER \citep{weaver1978presupernova,rauscher2003hydrostatic}.  Here we focus on models of primordial stars with initial compositions corresponding to Big Bang nucleosynthesis from \citet{cyburt2002}.  We explore models of mass ranging from $20$--$35\,\Msun$, with initial rotation speeds $\vrot$, ranging from $40$--$70\,\%$ of critical speed, $\vcrit$.  We name the models according to progenitor metallicity, mass, $\vrot/\vcrit$, and the mass loss prescription. For example, a zero metallicity star of $25\,\Msun$ with $\vrot/\vcrit=60\,\%$ and mass loss model ${\rm WR_0}$ is named \texttt{z25WR$_0$60}.
We use a co-processing adaptive reaction network to calculate multizone nucleosynthesis from the birth of a star to its death via CCSN with reaction rates as detailed in~\citet{rauscher2002nucleosynthesis}.

The details of the implementation of rotation in \KEPLER are discussed in~\citet{heger2000presupernova}.  In brief, the effects of various rotation-induced instabilities such as Eddington-Sweet circulation, dynamical shear instability, Solberg-H{\o}iland instability, secular shear instability,  and Goldreich-Schubert-Fricke instability are taken into consideration.   The resulting mixing induced by rotation is modelled as an additional diffusion coefficient.  The efficiency of rotational mixing is controlled by two free parameters, \fc, and \fmu.  The former is an overall multiplicative factor to the diffusion coefficient resulting from rotation-induced mixing and the latter controls the sensitivity to the compositional gradient. 
We adopt fiducial values of \fc\, and \fmu\, of $\nicefrac1{30}$ and $0.05$, respectively.  These values provide a good fit to the enrichment of surface nitrogen abundance observed in massive rotating stars of $10$--$20\,\Msun$ at solar metallicity~\citep{gies1992,villamariz2005,heger2000presupernova}. 
Magnetic fields resulting from the Taylor–Spruit dynamo ~\citep{spruit2002dynamo} are implemented including the effect of magnetic torque on angular momentum transport as discussed in~\citet{heger2005presupernova}.

\subsection{Mass loss}
Mass loss is crucial in rapidly rotating stars.   For progenitors that have not reached the Wolf-Rayet (WR) phase, we use the mass loss rates from \citet{nieuwenhuijzen1990parametrization}.  This rate is essentially zero for metal-free progenitors.  When the surface H mass fraction drops below $40\,\%$ and the surface temperature exceeds $\Ep4\,\K$, a star is considered to have entered the WR phase.  
The default mass loss rate prescription for the WR stage is adopted from \citet{yoon2006single}.  The rate is adapted from \citet{hamann1995spectral} that is reduced by a factor of $10$.  This is consistent with the mass loss rate reported in \citet{vink2005metallicity}.  We refer to this rate as $\rm{WR_0}$.

Metallicity plays a crucial role in mass loss as the strength of radiative winds directly impacts the overall rate of mass loss.  \citet{vink2005metallicity} found that for WR stars the mass loss rate scales with effective metallicity $Z_\mathrm{eff}=Z_{\mathrm{Fe,surf}}/Z_\mathrm{Fe,\odot}$ as $\approx Z_{\mathrm{ eff}}^{0.86}$, where $Z_{\rm Fe,surf}$ is the surface Fe mass fraction and $Z_{\mathrm{Fe},\odot}$ is the mass fraction of Fe in the Sun. Thus, with this metallicity dependence the mass loss rate $\dot{M}$ for ${\rm WR_0}$ is given by 

\begin{align}
 \label{eq:mlossWR}
  &\log\left(\frac{\dot{M_{\rm WR_0}}}{\Msun \,\yr^{-1}}\right)\nonumber\\
  &\qquad=
    -12.95+1.5 \,\log \left (\frac{L}{\Lsun}\right)-2.85 X_{\mathrm{H,surf}}+0.86 \,\log \left({Z_{{\rm eff}}}\right), \nonumber\\
    & \qquad\qquad\qquad\qquad\qquad\qquad\qquad\text{for}\; \log \left(\frac{L}{\Lsun}\right) > 4.5  \nonumber\\
    &\qquad= -36.8+6.8 \,\log \left(\frac{L}{\Lsun}\right)+0.86 \,\log \left(Z_{{\rm eff}}\right),  \\
    & \qquad\qquad\qquad\qquad\qquad\qquad\qquad\text{for}\; \log \left(\frac{L}{\Lsun}\right) \leq 4.5  \nonumber
\end{align}
where $L$ is the surface luminosity of the star and $X_{\mathrm{H,surf}}$ is the surface $^1$H mass fraction.

Although the metallicity dependence on $Z_{\rm Fe}$ is valid for a wide range of metallicities, 
\citet{vink2005metallicity} found a particularly interesting feature for extremely metal-poor WR progenitors. They found that the dependency of mass loss rate on $Z_{\rm Fe,surf}$ is no longer valid for WR stars that have an initial metallicity of $Z_{\rm Fe}/Z_{\rm Fe,\odot}\lesssim\Ep{-4}$ when their surface is enriched in N (WN stars) or C (WC stars).  In such extremely metal-poor WN or WC stars, the winds are radiatively driven by intermediate elements such as C and N rather than heavier elements like Fe and the mass loss rate is \textit{effectively independent of} $Z_{\rm Fe,surf}$.  This implies that massive stars of very low initial metallicity including metal-free progenitors can undergo mass loss once they reach the WN or WC phases. 

In order to implement this behaviour of mass loss rate for WN stars we modify the mass loss rate formula in Eq.~\ref{eq:mlossWR} for $Z_{\rm Fe,surf}/Z_{\rm Fe,\odot}\leq\Ep{-4}$ by modifying $Z_{\rm eff}$
as follows
\begin{equation}
    Z_{\rm eff}= {\rm min }\left[\Ep{-4},\frac{Z_{\rm Fe,surf}}{Z_{\rm Fe,\odot}} + \Ep{-4}\frac{X_{\rm CNO,surf}}{X_{\rm CNO, \odot}}\right]
\end{equation}
where $X_{\rm CNO,surf}$ is the total surface mass fraction of CNO and $X_{\rm CNO,\odot}$ is the corresponding value in the Sun. 
The above formula results in $Z_{\rm eff}=\Ep{-4}$ when $X_{\rm CNO,surf}$ exceeds $X_{\rm CNO,\odot}\approx0.01$.
\citet{vink2005metallicity} also found that for extremely metal-poor WC stars that have very high surface enrichment of C (surface mass fraction of C$\gtrsim0.1$), the mass loss rate is higher compared to WN stars by up to a factor of $\lambda\approx10$.  In order to implement this, the mass loss rate for WC stars is taken to be 
\begin{equation}
 \label{eq:mloss1_wc}
    \dot{M}_{\rm WC} =\eta_{\rm WC}\,\dot{M}_{\rm WR} \qquad\text{for} X_{\rm C,surf}> X_{\rm N,surf}
\end{equation}
where
\begin{equation}
  \label{eq:mloss2_wc}
    \eta_{\rm WC}=1 +{\rm min}[(\lambda-1),10\,X_{\rm C,surf}\,(\lambda-1)],
\end{equation}
and $X_{\rm C,surf}$ and $X_{\rm N,surf}$ are the surface mass fraction of  C and N, respectively.
The above formulation ensures that $\eta_{\rm WC}$ increases smoothly from 1 up to a maximum value of $\lambda=10$ as the $X_{\rm C,surf}$ reaches a value of $0.1$.

We also model the enhancement of mass loss rate due to stellar rotation based on the rate from \citet{Langer1997ASPC, yoon2005evolution} given by
\begin{equation}
    \label{eq:lossrot}
    \dot{M}=\dot{M}(\vrot=0)\cdot\left(\frac{1}{1-v/\vcrit}\right)^{0.43},
\end{equation}
where $\dot{M}(v_{\rm rot}=0)$ corresponds to the mass loss rate for the non-rotating model and $v_{\rm crit}$ is the critical velocity given by 
\begin{equation}
    \vcrit=\sqrt{\frac{GM}{R}(1-\Gamma)}
\end{equation}
where $M$ and $R$ are the mass and radius of the star, respectively.  $\Gamma$ is the Eddington factor given by
\begin{equation}
    \Gamma=\frac{\kappa L}{4\pi cGM}
\end{equation}
where $\kappa$ is the opacity at the surface.  In rapidly-rotating models, the value of $v/\vcrit$ starts to approach unity following central H depletion which would result in the mass loss rate blowing up as per Eq.~\ref{eq:lossrot}.  In order to avoid this, the maximum value of $v/\vcrit$ in Eq.~\ref{eq:lossrot} is limited to a value of $0.99$.  This corresponds to a maximum rotation-induced enhancement factor of $\sim7.2$.

\section{Results}\label{sec:evolution and results}

We find that for stars of all initial masses, when the $\vrot$ is above a certain fraction of  $\vcrit$,  rotationally-induced mixing leads to the formation of a He star that is nearly chemically homogeneous at core hydrogen depletion i.e., a QCH star.  This is consistent with the earlier rapidly-rotating models of \citet{yoon2005,woosley2006progenitor} and, more recently, of \citet{banerjee2019}.
The minimum ratio of initial $\vrot/\vcrit$ required to reach the QCH stage decreases as the mass of the progenitor increases.
 For example, for a star with $20\,\Msun$ initial mass, the minimum rotation speed required to reach the QCH state is a $\vrot$ of $49\,\%$ of $\vcrit$, but only $40\,\%$ for a $35\,\Msun$ star which correspond to $45\,\%$ and $35\,\%$ of break-up speed, respectively.  Below we discuss the details of the evolution of models that reach QCH state using the \texttt{z25WR$_0$60} model as a fiducial model.

\subsection{Evolution of QCH stars}

\begin{figure*}
    \centering
    \includegraphics[width=\textwidth]{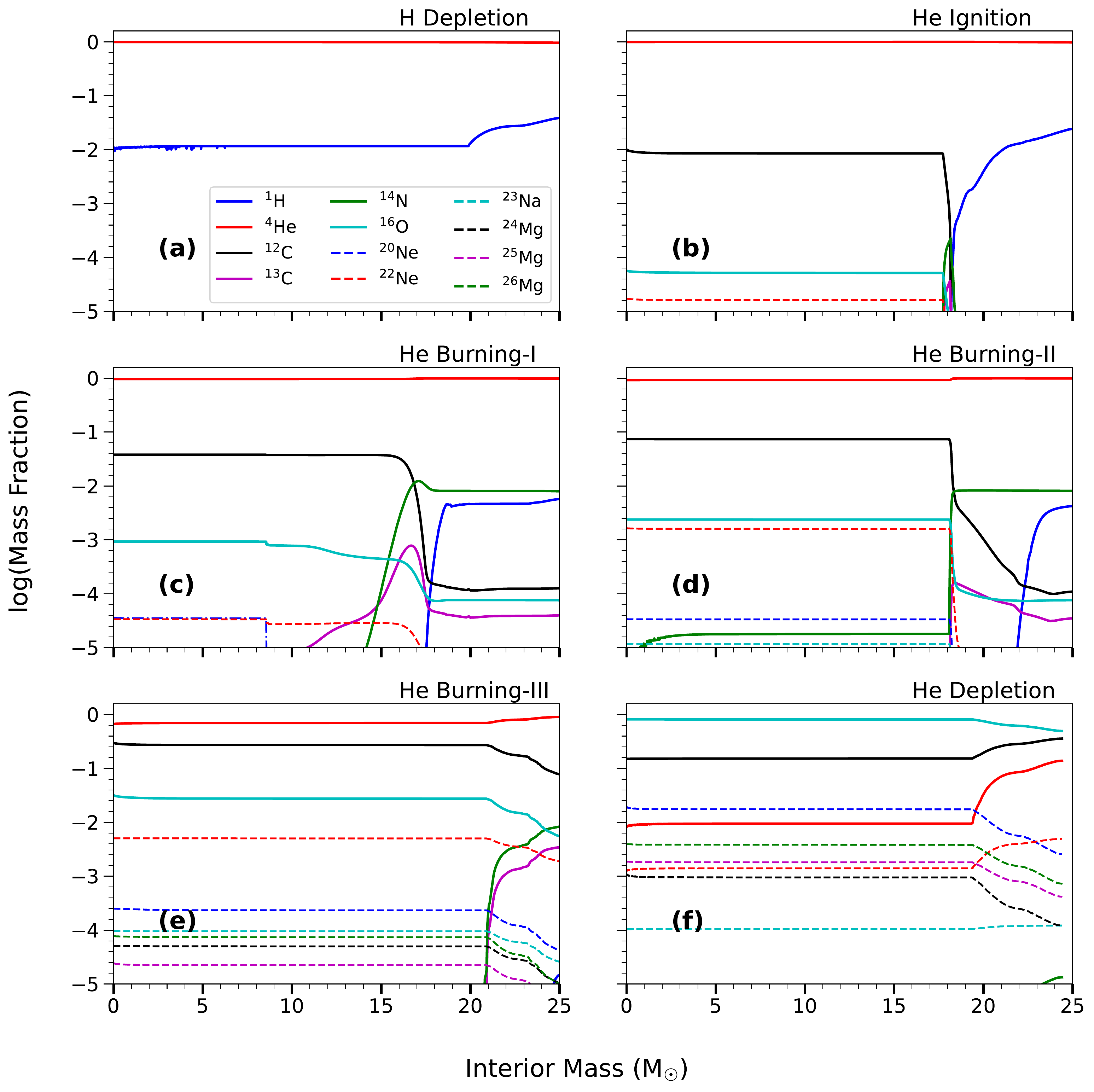}
    \caption{Mass fractions of light isotopes as a function of mass coordinate during different stages of core He burning of the  \texttt{z25WR$_0$60} model.}
    \label{fig:z25_evolve}
\end{figure*}

\begin{figure*}
    \centering
    \includegraphics[width=\textwidth]{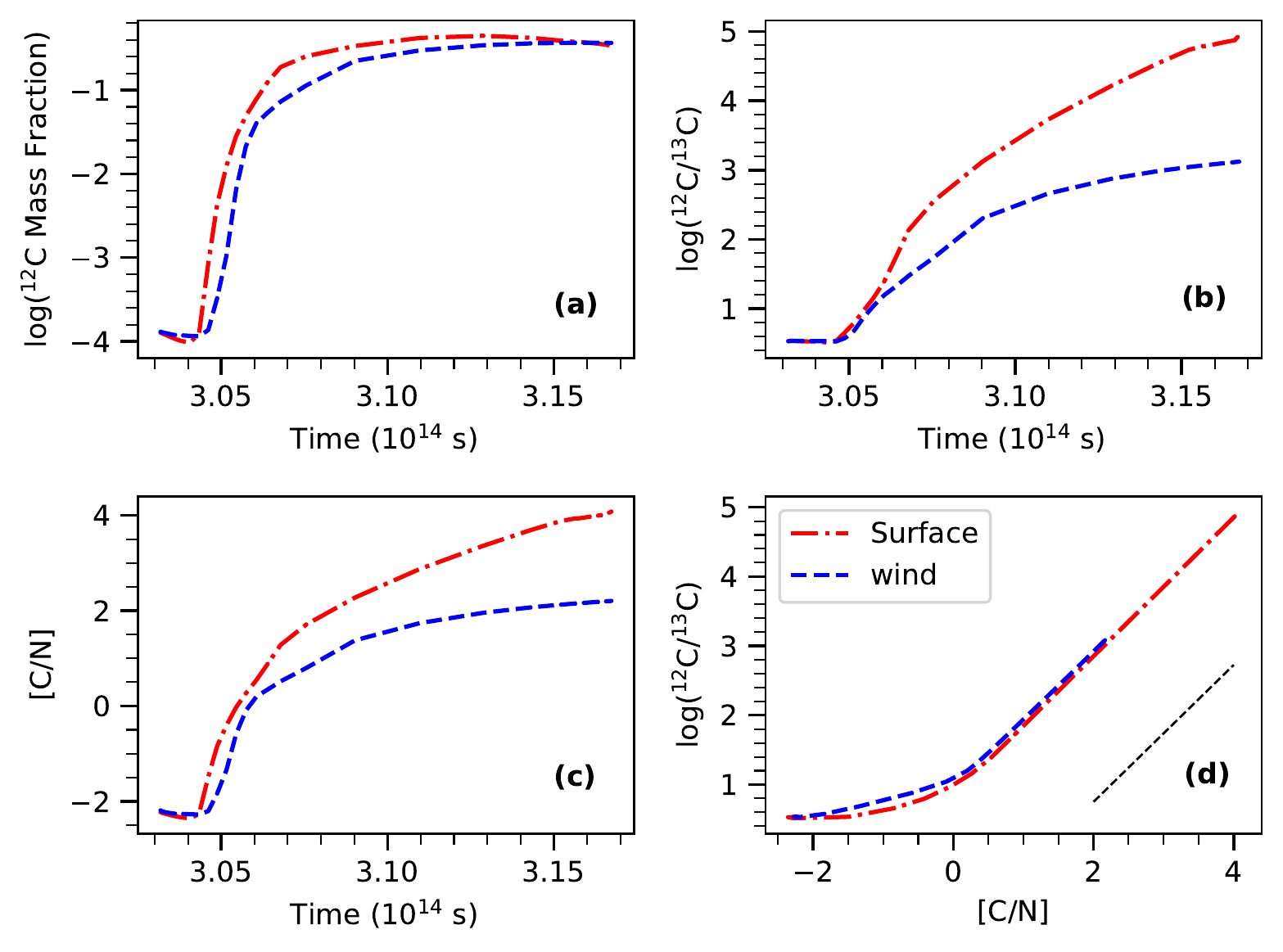}
    \caption{The time evolution of $^{12}$C (Panel~a), $^{12}$C/$^{13}$C (Panel~b) and $\B{C}{N}$ (Panel~c) at the surface (optical depth $\tau=\nicefrac23$) as well as in the cumulative wind ejecta of  \texttt{z25WR$_0$60} model.  The \textsl{red dash-dotted line} and the \textsl{blue dashed line} show the values at the surface and in the total wind ejecta, respectively.  Panel~(d) shows the corresponding evolution of  $^{12}$C/$^{13}$C with $\B{C}{N}$ where the \textsl{black dashed inclined line} in Panel~(d) provides a reference line with a slope of 1. }
    \label{fig:cno_z2560}
\end{figure*}

Figure~\ref{fig:z25_evolve} shows the evolution of mass fractions of most abundant isotopes at different stages of evolution for the  \texttt{z25WR$_0$60} model from the core H depletion (Fig.~\ref{fig:z25_evolve}a) stage to the core He depletion stage (Fig.~\ref{fig:z25_evolve}f).  The different stages are shown in Fig.~\ref{fig:z25_evolve} as Panels a--f.  The stages are defined as follows:
\begin{itemize}

    \item[(a)] H Depletion: when the central $^1$H mass fraction drops to $1\,\%$.
    \item[(b)] He Ignition: when the central $^{12}$C mass fraction reaches $1\,\%$.
    \item[(c)] He Burning-I: when the surface $X_{\rm CNO}$ which is dominated by $X_{\rm N,surf}$, reaches $X_{\rm CNO, \odot}$.
    \item[(d)]  He Burning-II: when H is exhausted in the inner part of the non-He burning core.
    \item[(e)]  He Burning-III: when the central $^4$He mass fraction drops to $50\,\%$
    \item[(f)]  He Depletion: when the central $^4$He mass fraction drops to $1\,\%$.
\end{itemize}

During central H burning, the convective core continues to grow as rotation-induced mixing results in protons from the envelope being mixed into the core. As a result, by the time the central H mass fraction drops to $\sim30\,\%$, the star enters the WR phase as the surface H mass fraction drops below $40\,\%$ and an effective temperature of $\gtrsim\Ep5\,\K$.
By the time the star reaches the H Depletion stage, the He burning core (HeBC) covers $\gtrsim80\,\%$ of the star by mass.  The outer non-He burning part of the core (non-HeBC) which is non-convective extends up to $\gtrsim99\,\%$ of the star, by mass, with a negligible envelope mass.  At this stage, almost the entire star is composed of $^4$He, including the outer core, which has a $^1$H mass fraction of $\sim 1\,\%$--$3\,\%$, and the star has reached the QCH stage (Fig.~\ref{fig:z25_evolve}a).

As $^4$He starts burning in the center, $^{12}$C and $^{16}$O are synthesized in the convective HeBC.  Due to rotation-induced mixing, the elements synthesized inside the HeBC are slowly mixed with the non-HeBC where a small mass fraction of $^1$H is still present. 
Because the QCH star is essentially a He star, there is effectively no hydrogen envelope and the temperature and density gradients are shallow outside the HeBC.
The typical values just outside the HeBC at the He Ignition stage are $\sim7$--$8\E7\,\K$ and $\sim 30~\g\,\cm^{-3}$, respectively.  As a result, $^{12}$C in the outer core is converted into $^{14}$N via the CNO cycle.  This material is then mixed out all the way to the surface layers due to rotation-induced mixing.  Consequently, the surface CNO abundance, which is dominated by the $^{14}$N abundance, gradually increases.  Even though the progenitor is metal-free, this marks the start of mass loss as per Eq.~\ref{eq:mlossWR} where the effective metallicity $Z_{\rm eff}$ quickly reaches the maximum value of $10^{-4}$ as the surface $X_{\rm N}$ reaches $\gtrsim1\,\%$ by the He Burning-I stage (Fig.~\ref{fig:z25_evolve}c). 
Following this stage, H in the outer core is entirely consumed by the CNO process.  As a result, the surface $X_{\rm C}$ starts to increase as an increasing amount of $^{12}$C is mixed out of the convective HeBC which is no longer processed by the CNO cycle.  Eventually, the surface value of  $X_{\rm C}$ exceeds $X_{\rm N}$ and reaches values of $\gtrsim10\,\%$ as the star enters the WC stage. 
This results in an increase in mass loss rate  by a factor of up to $10$ (Eqs.~\ref{eq:mloss1_wc} and \ref{eq:mloss2_wc}).  The majority of the mass loss occurs post this stage which lasts up to $\gtrsim300\,\kyr$ and accounts for $\gtrsim90\,\%$ of the total mass loss of $\sim0.63\,\Msun$ due to stellar wind.

\subsection{Time-dependent wind composition}

As the material from the inner part of non-HeBC is gradually mixed out and reaches the surface, the surface composition changes over time and reflects the composition of the inner non-HeBC at an earlier time.  
The composition of the wind ejecta changes over time accordingly.  Figure.~\ref{fig:cno_z2560} shows the details of the evolution of the CNO abundances at the surface, as well as the cumulative wind ejecta.
The total wind ejecta shown for a given instant is the cumulative average of the past surface compositions weighed by the mass loss rate. 

In our Pop III stars, mass loss only sets in when mixing brings CNO-enriched material to the surface for the first time.  Following this, the evolution of the surface composition can be divided broadly into the following three phases:
\begin{enumerate}

    \item During the initial phase, as $^{12}$C from the HeBC starts to mix out into the non-HeBC due to rotation-induced mixing, it is quickly converted into $^{14}$N via the CNO cycle as the abundance of $^1{\rm H}\sim 1\,\%$ is sufficiently high.  This results in CNO abundance ratio\footnote{defined as number fraction ratio} at the surface that correspond to the CNO equilibrium values of $^{12}$C/$^{13}$C of $\sim 4$ and $\B{C}{N}\sim-2$.  He Burning-I (Fig.~\ref{fig:z25_evolve}c) belongs to this phase.  This phase lasts for $\sim30\,\kyr$, and $\lesssim0.005\,\Msun$ are lost due to stellar wind during this phase.

    \item During the next phase, as more $^{12}$C is mixed into the non-HeBC and the $^1$H abundance starts to get depleted in the inner non-HeBC.  As the number abundance of $^1$H becomes lower than that of $^{12}$C, $^{12}$C is only partially processed via the CNO cycle such that $^{12}$C/$^{13}$C remains roughly unchanged whereas $\B{C}{N}$ in the inner non-HeBC increases by more than an order of magnitude to $\gtrsim -1$.  The surface CNO composition, however, remains largely unchanged and still corresponds to CNO equilibrium abundance values as in the initial phase.  During this phase, the $^{14}$N synthesized in the non-HeBC gradually starts to mix back into the convective HeBC where it is efficiently converted to $^{22}$Ne as it reaches the center where the temperature is $\sim2\E8\,\K$.  He Burning-II (Fig.~\ref{fig:z25_evolve}d) belongs to this phase.  This phase lasts for $\lesssim30\,\kyr$ and $\lesssim0.002\,\Msun$ are lost due to stellar wind.

    \item Finally, as $^1$H is completely exhausted in the non-HeBC, the increasing amount of $^{12}$C that is mixed out from the HeBC is completely unprocessed by the CNO cycle and reaches the surface layers.  As a result, the surface  mass fraction of $^{12}$C is increased by $\sim2$ orders of magnitude and the star becomes a WC star.  The value $^{12}$C/$^{13}$C and C/N also increase by the same amount leading to a linear slope of $^{12}$C/$^{13}$C vs C/N.  This can be clearly seen from Fig.~\ref{fig:cno_z2560}d where the slope  of the curve for $^{12}$C/$^{13}$C vs $\B{C}{N}$ is $\sim1$ beyond the He Burning-III stage.  This phase is the longest of the three phases which lasts for $\gtrsim300\,\kyr$ and accounts for most of the mass loss. During the initial stages of this phase, $^{22}$Ne, which is synthesized in the HeBC, also mixes out into the non-HeBC and reaches the surface layers. During this phase, $^{23}$Na is also synthesized in the core via neutron capture on $^{22}$Ne where the neutrons are created via $^{13}$C($\alpha$,n)$^{16}$O and also mixes out to the surface. Subsequently, as the central core temperature reaches values $\gtrsim2.5\E8\,\K$, most of the $^{22}$Ne in the HeBC is burned into $^{26}$Mg (and $^{25}$Mg) whereas the $\alpha$ capture on $^{16}$O results in a minor production of $^{20}$Ne and $^{24}$Mg.  These isotopes again are mixed out to the non-HeBC and eventually reach the surface layers. 
    He Burning-III (Fig.~\ref{fig:z25_evolve}e) and He Depletion (Fig.~\ref{fig:z25_evolve}f) correspond to the initial and latter stages of this phase. 
\end{enumerate}

\subsection{Evolution of wind ejecta}
\begin{figure}
    \centering
    \includegraphics[width=\columnwidth]{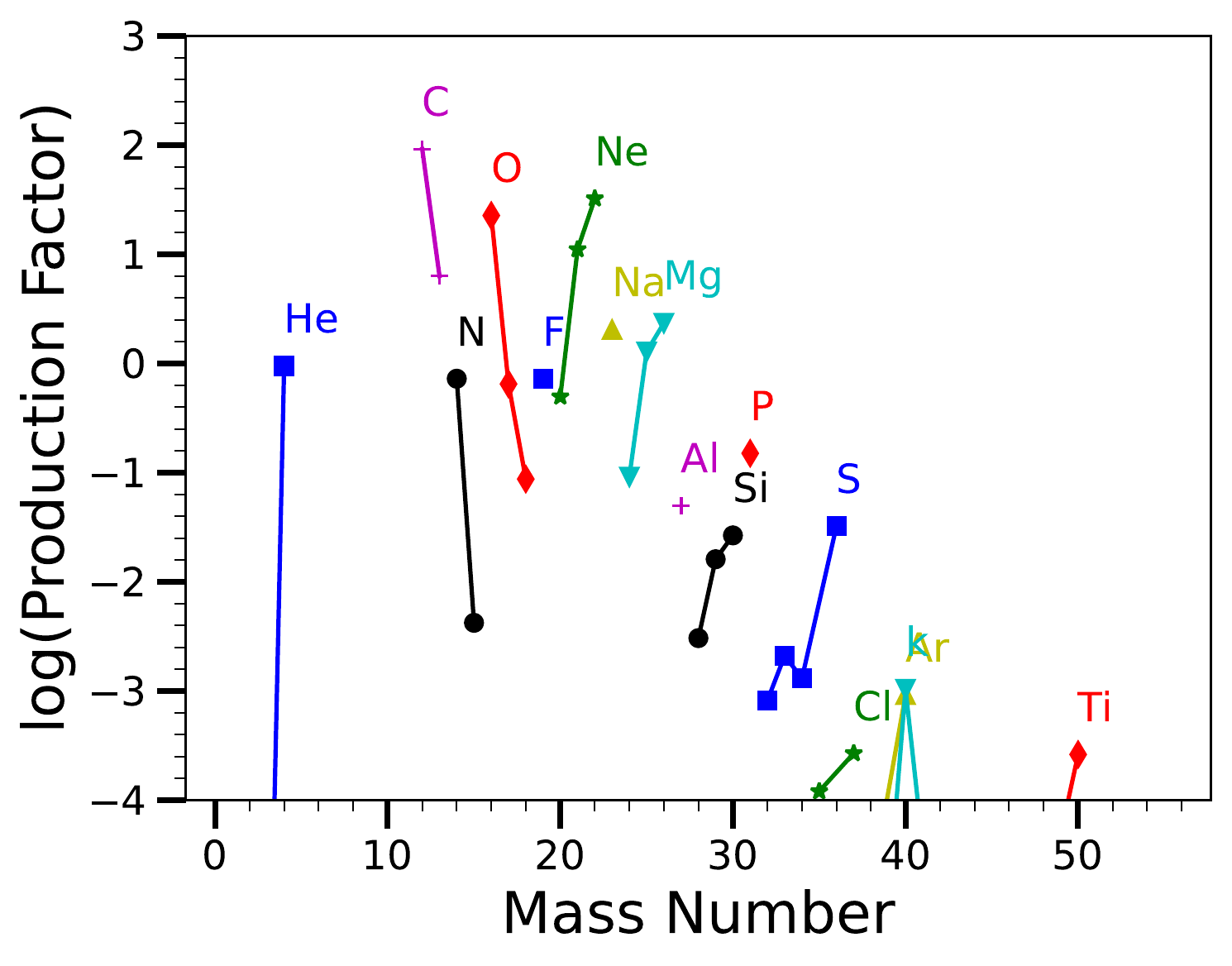}
    \caption{Isotopic abundance pattern relative to the solar value of the total wind ejecta for the  \texttt{z25WR$_0$60} model.}
    \label{fig:wind_isotopic}
\end{figure}

As the surface CNO abundance changes during the three phases described above, the cumulative wind composition also changes gradually. 
Figure~\ref{fig:cno_z2560}c shows that the wind starts out as N-rich during the initial phase 
but quickly becomes C-rich with $\B{C}{N}\gtrsim 0$ when it enters the final phase where most of the mass is lost.  During this time, because the surface composition corresponds to material that has undergone partial processing, the $^{12}$C/$^{13}$C remains $\lesssim 10$ even as $\B{C}{N}$ increases to values above 0.  Beyond this point, unprocessed $^{12}$C is lost from the surface such that, by the end of the final phase, the cumulative wind composition becomes extremely enhanced in C (and O) with $\B{C}{N}\sim 2$ and $^{12}$C/$^{13}$C$\sim 1000$.  During the last stages of the final phase (WC phase), isotopes up to $^{26}$Mg that have reached the surface are also incorporated in the wind. The final wind ejecta is essentially dominated by $^{4}$He ($41\,\%$), $^{12}$C ($37\,\%$), and $^{16}$O ($22\,\%$) where the absolute abundance of C is higher than O with $\B{C}{O}= 0.6$. Figure.~\ref{fig:wind_isotopic} shows the production factor of the final wind ejecta.  The figure shows that in addition to CNO, a significant amount of F, Na, Ne, and Mg isotopes is also produced.  Interestingly, among the Mg isotopes, the abundance of $^{25}$Mg and $^{26}$Mg is higher than that of $^{24}$Mg (Table~\ref{tab:z25_rotation}). Similarly, the abundance of $^{22}$Ne is higher than that of $^{20}$Ne.  This is very different from typical core-collapse SN yields in which the Ne and Mg yields are almost entirely dominated by $^{20}$Ne and $^{24}$Mg.

\subsection{Effect of rotation rate and progenitor mass on wind ejecta} \label{effectofrotation_on_QCH}
In order to explore the effect of rotation rate on mass loss and wind ejecta, we calculated $25\,\Msun$ models with varying rotation speeds ranging from ${\rm v_{rot}}/v_{\rm crit}=0.46$ to $0.70$, where the lowest rotation rate corresponds to the minimum initial rotation required to reach the QCH state.  Table~\ref{tab:z25_rotation} lists the major isotopes that comprise the winds ejecta. For ${\rm v_{rot}}/v_{\rm crit}\gtrsim 0.50$, the amount of mass lost in the wind as well as the yields of major isotopes $^{4}$He, $^{12}$C, and $^{16}$O are approximately proportional to the rotation rate.  The increase in $^{12}$C and $^{16}$O reflect the higher mixing efficiency in the late stages of He burning, resulting in higher enrichment of these isotopes on the surface.  The yield of  $^{14}$N, on the other hand, decreases with increasing rotation rate.
As the rotation rate increases, a lower amount of $^{1}$H is left in the non-HeBC after central H depletion. The protons in the non-HeBC are subsequently converted into $^{14}$N as $^{12}$C mixes out of the HeBC. Thus, models with higher rotation have lower $^{14}$N abundance in the non-HeBC as well as at the surface zones.  For example, during WC phase, when almost all of the mass loss occurs, on average, $X_{\rm N,surf}$ for Model \texttt{z25WR$_0$46} is a factor of $\sim 3$ higher as compared to Model \texttt{z25WR$_0$60}.  Although $M_{\rm wind}$ is lower in Model \texttt{z25WR$_0$46} by a factor of $\sim 1.5$, the higher value of $X_{\rm N,surf}$ leads to a factor $\sim 2$ higher absolute yield of $^{14}$N in the wind compared to Model \texttt{z25WR$_0$60}.  The abundance of $^{14}$N relative to $^{12}$C is a factor of $\sim 3$ higher in Model \texttt{z25WR$_0$46} compared to Model \texttt{z25WR$_0$60} since the former has a lower $^{12}$C yield due to slower rotation.

\begin{figure}
    \centering
    \includegraphics[width=\columnwidth]{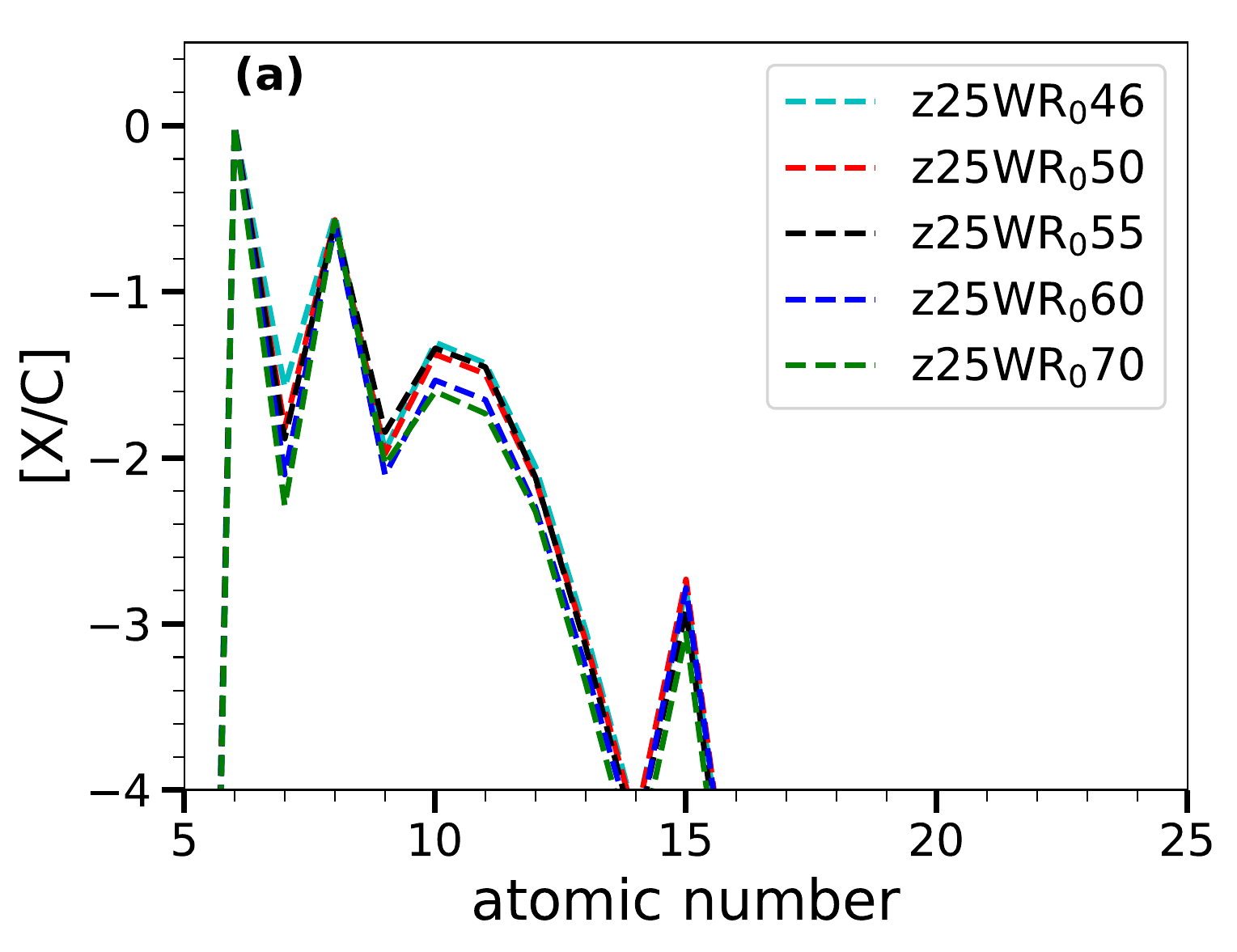}
    \includegraphics[width=\columnwidth]{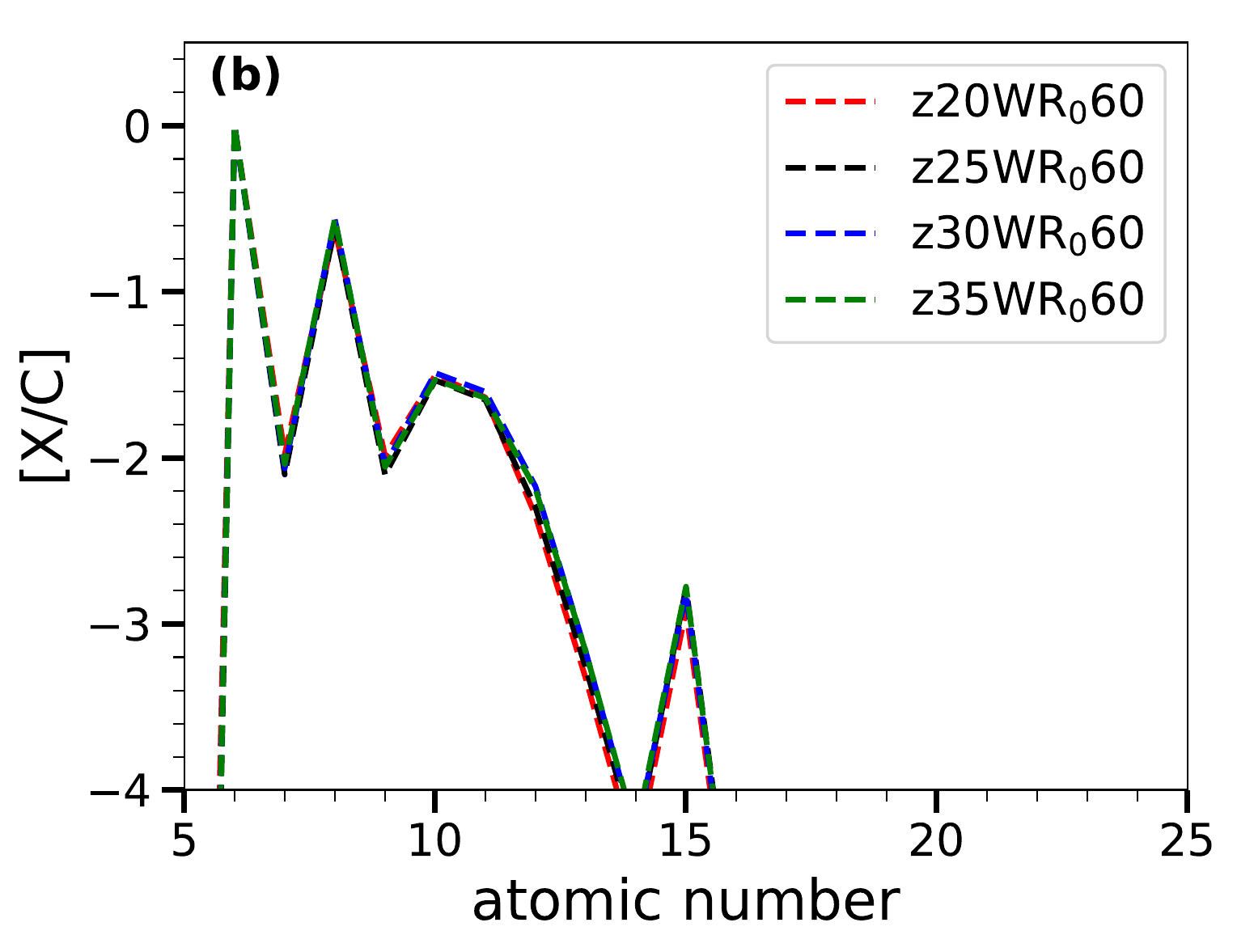}
    \caption{Elemental abundance pattern relative to C produced by the total wind ejecta of various rapidly rotating massive models.  \textsl{Panel~(a)}: Elemental pattern for the z25$\Msun$ model with varying values of $v_{\rm rot}/v_{\rm crit}$ of $0.46$--$0.7$.  \textsl{Panel~(b)}: Elemental pattern corresponding to models of different masses  with the same $v_{\rm rot}/v_{\rm crit}$ of $0.6$.}
    \label{fig:wind_pattern}
\end{figure}

The trend of monotonically increasing yields of $^{12}$C and $^{16}$O and decreasing yield of $^{14}$N with increasing rotation rate are not seen for F, Na, Ne, and Mg.  This is because the synthesis of these elements depends primarily on the amount of $^{14}$N that is mixed from the outer non-HeBC into the HeBC where the isotopes of these elements are synthesized and eventually mixed out to the surface.  Whereas higher $^{14}$N is available in the non-HeBC core for slower rotation, the rate of mixing of $^{14}$N into the HeBC as well as the rate of material mixed out of the HeBC to the surface is lower and vice-versa.  Thus, the yields of F, Na, Ne, and Mg in the wind initially increase with the rotation rate and reach a maximum at an optimum rotation rate. For the \texttt{z25WR$_0$} models, the maximum yield occurs at a rotation rate of $55\,\%$ of $v_{\rm crit}$ (Table \ref{tab:z25_rotation}).  
The elemental abundance patterns relative to C for the various \texttt{z25} models are shown in Fig.~\ref{fig:wind_pattern}a.  The relative abundances for C and O are almost identical with a value of $\B{C}{O}\sim 0.6$ across all models.  On the other hand, relative to C, the abundance of N varies by a factor of $\sim 5$, whereas the abundances of F, Na, Ne, and Mg vary by a factor of $\sim 2$ as $v{\rm_{rot}}$ increases from $46\,\%$ to $70\,\%$ of $v_{\rm crit}$.

We also explored the effect of progenitor mass on the properties of wind ejecta.  In Table~\ref{tab:progenitor_mass}  we list the wind ejecta yields for progenitors of mass $20\,\Msun$--$35\,\Msun$ with ${\rm v_{rot}}/v_{\rm crit}=0.60$.  Overall, the total mass lost in the wind increases with progenitor mass with a corresponding increase in the yields of the isotopes. The relative elemental abundance pattern is almost identical for all progenitors (Fig.~\ref{fig:wind_pattern}b). 

\begin{table*}
 \centering
  \caption{The yields of most abundant elements and total mass ejected in the wind $M_{\rm {wind}}$ in units of $\Msun$ in total wind ejecta for $25\,\Msun$ progenitors with initial rotation speeds ranging from ${\rm v_{rot}}/v_{\rm crit}=0.46$ to $0.70$. Note that $X(Y)\equiv X\times10^Y$.}
  \resizebox{\textwidth}{!}{%
\begin{tabular}{ccccccccccccccc}
    \hline\hline
    Model&$^{4}$He &$^{12}$C &$^{13}$C &$^{14}$N &$^{16}$O & $^{19}$F& $^{20}$Ne &$^{22}$Ne&$^{23}$Na &$^{24}$Mg&$^{25}$Mg &$^{26}$Mg &$M_{\rm {wind}}$&$\nicefrac{M_{\rm C}}{M_{\rm wind}}$\\

       \hline\hline
        \texttt{z25WR$_0$46}&1.56(-1)& 1.45(-1)&3.08(-4)&1.12(-3)&1.01(-1)&2.45(-7) &6.06(-4)&3.43(-3)&7.07(-5)&4.54(-5)&9.67(-5)&1.97(-4)&0.41&0.35\\
        \texttt{z25WR$_0$50}&1.91(-1)&1.87(-1)&2.47(-4)&8.25(-4)&1.22(-1)& 3.04(-7)
 &6.61(-4)&3.76(-3)&7.87(-5)&5.14(-5)&1.02(-4)&2.13(-4)&0.51&0.37\\
        \texttt{z25WR$_0$55}&2.26(-1)&2.18(-1)&2.00(-4)&8.29(-4)&1.36(-1)& 4.77(-7)
 &6.63(-4)&4.96(-3)&1.00(-4)&6.22(-5)&1.28(-4)&2.49(-4)&0.59&0.37\\  
       \texttt{z25WR$_0$60}&2.57(-1)&2.31(-1)&1.89(-4)&5.34(-4)&1.38(-1)&2.80(-7) &6.14(-4)&3.19(-3)&6.78(-5)& 4.61(-5)&8.45(-5)&1.76(-4)&0.63&0.37\\
       
       \texttt{z25WR$_0$70}& 3.18(-1)&2.62(-1)&1.15(-4)&3.93(-4)&1.69(-1)& 3.65(-7)
 &6.73(-4)&3.01(-3)&6.35(-5)&5.04(-5)&9.84(-5)&1.81(-4)&0.75&0.35\\
       
       \hline   
    \end{tabular}}
   
    \label{tab:z25_rotation}
\end{table*}

\begin{table*}
    \centering
     \caption{Same as Table~\ref{tab:z25_rotation} but for varying progenitor mass  20--${\rm 35\,\Msun}$ with same initial rotation speed of ${\rm v_{rot}}/v_{\rm crit}=0.60$. }
     \resizebox{\textwidth}{!}{%
    \begin{tabular}{ccccccccccccccc}
    \hline\hline
     Model&$^{4}$He &$^{12}$C &$^{13}$C &$^{14}$N &$^{16}$O& $^{19}$F & $^{20}$Ne & $^{22}$Ne&$^{23}$Na &$^{24}$Mg&$^{25}$Mg &$^{26}$Mg &$M_{\rm {wind}}$&$\nicefrac{M_{\rm C}}{M_{\rm wind}}$\\

       \hline\hline
       
        \texttt{z20WR$_0$60}&1.60(-1)&1.45(-1)&1.38(-4)&4.43(-4)&8.33(-2)& 2.29(-7)
 &3.78(-4)&2.18(-3)&4.40(-5)&2.82(-5)&4.64(-5)&9.71(-5)&0.39&0.37\\
       \texttt{z25WR$_0$60}&2.57(-1)&2.31(-1)&1.89(-4)&5.34(-4)&1.38(-1)& 2.80(-7)
 &6.14(-4)&3.19(-3)&6.78(-5)& 4.61(-5)&8.45(-5)&1.76(-4)&0.63&0.36\\
       
        \texttt{z30WR$_0$60}&3.88(-1)&3.36(-1)&2.55(-4))&8.48(-4)&2.21(-1)& 4.65(-7)
 &1.03(-3)&5.10(-3)&1.10(-4)&8.21(-5)&1.75(-5)&3.41(-4)&0.95&0.35\\
        \texttt{z35WR$_0$60}&5.51(-1)&4.37(-1)&4.08(-4)&1.17(-3)&2.90(-1)& 5.89(-7)
 &1.31(-3)&5.92(-3)&1.31(-4)&9.87(-5)&2.19(-4)&4.33(-4)&1.29&0.34\\

       \hline   
    \end{tabular}}
   
    \label{tab:progenitor_mass}
\end{table*}

\subsection{Evolution following core-collapse}
Since stars that reach the QCH stage evolve essentially as a He star, the core constitutes $\gtrsim 95\,\%$ of the star by mass.  The size of the core thus corresponds to the core of a non-rotating star of much larger mass \citep{banerjee2019}. The structure of such stars is extremely compact when the star enters the core collapse phase.  This can be quantified by the compactness parameter $\xi_M$  defined as~\citep{o2011black,muller2016simple}
\begin{equation}
    \xi_M=\left(\frac{M}{\Msun}\right)  \left( \frac{1000\,\rm km}{R(M)}\right),
    \label{eq:4}
\end{equation}
where $M$ is the mass of the inner core and $R(M)$ is the corresponding radius.  We evaluate $\xi_M$ at the pre-SN stage, which we define as the instant when the infall velocity exceeds $900~{\rm km\,s^{-1}}$ \citep{WZW95}.
Similar to~\citet{o2011black}, we adopt $M=2.5\,\Msun$ as it is relevant for the typical mass scale for black hole formation.  It has been shown that the value of $\xi_{2.5}$ can be used to predict whether a star undergoes a successful explosion by the neutrino-driven mechanism or collapses into a black hole following the core collapse~\citep{muller2016simple}.  The values of $\xi_{2.5}$ for the various rapidly rotating massive star models range from $0.34$--$0.80$.  These are much higher than the maximum value of $\sim 0.28$ reported by \citet{muller2016simple}, above which stars are not expected to undergo successful neutrino-driven explosion.  Additionally, when the two-parameter criterion for estimating the explodability by \citet{ertl2016two} is applied to our models, all of them lie in the region of parameter space where the models do not explode.  Thus, we expect that the QCH models are unlikely to explode via the usual neutrino-driven explosion mechanism. 

The above estimates of explodability, however, may not directly apply to rapidly rotating models.  Results from detailed three-dimensional magnetohydrodynamic simulations with neutrino transport suggest that some rapidly-rotating models could explode via a rotationally aided neutrino-driven mechanism while others may not \citep[e.g.,][]{mosta2014,kuroda2020,powell2020}.  On the other hand, if sufficiently strong  magnetic fields develop in rapidly rotating models, they can give rise to successful explosions via the magnetorotational mechanism with explosion energies ranging from weak to energetic SN explosions \citet{kuroda2020,grimmett2021,ober2021}.  Thus, it is likely that a substantial fraction of the QCH stars could explode as faint or regular SN whereas a certain fraction of them do not undergo any explosion.  When rapidly rotating QCH models do undergo successful explosion, because of their much larger cores that encompass almost the entire mass of the star, the SNcan result in the ejection of much larger amounts of intermediate elements, ranging from C to Ca, compared to non-rotating models of a similar initial mass.  In particular, the C yield for the QCH progenitors of $20$--$35\,\Msun$ explored here can eject a maximum C ranging from $\sim2$--$3\,\Msun$ compared to $\sim0.08$--$0.4\,\Msun$ for the non-rotating models of $12$--$30\,\Msun$. That is, rapidly rotating QCH models can potentially eject up to an order of magnitude more C than their non-rotating counterparts.

\section{Discussion} \label{sec:discussion}
Since some of the QCH models are unlikely to undergo successful explosion, we first explore the implications of the pollution of the ISM with the C-rich wind ejecta, in particular, for their potential to produce the abundance pattern of CEMP-no stars.  To estimate the mixing of the wind ejecta in the ISM, we approximate the wind as a blast wave of total energy equal to the total kinetic energy $E_{\rm wind}$ carried by the wind given by  
\begin{equation}
E_{\rm wind}=2.24 \times 10^{49}\left(\frac{M_{\rm wind}}{\Msun} \right) \left(\frac{v_{\rm wind}}{1500~{\rm km\,s^{-1}}} \right) ^{2}{\rm ergs}
      \label{eq:wind_K.E}
 \end{equation}
where $M_{\rm {wind}}$ is the total mass ejected in the wind and $v_{\rm wind}$ is the mass-average terminal velocity of the wind.  This is a reasonable approximation as most of the wind loss in our models happens in a relatively short interval during the WC phase that lasts for $\lesssim 300\,\kyr$.  We adopt a fiducial value of $v_{\rm wind}=1500\,{\rm km\,s^{-1}}$ that is consistent with the terminal velocities for optically thick winds for low-metallicity WR stars \citep{sander2020windA,sander2020windB}.  Using Eq.~(\ref{eq:wind_K.E}), we obtain $E_{\rm wind}=0.9$--$2.9\times10^{49}$~ergs for the various QCH models of with $M_{\rm wind}\sim 0.4$--$1.3\,{\rm M_\odot}$.  Thus, the winds from QCH models act as extremely weak explosions that have energies that are up to two orders of magnitude less than a regular CCSN.  

\subsection{Dilution of Wind Ejecta in Minihaloes} \label{sec:Dilution_minihalo}

\begin{figure}
    \centering
    \includegraphics[width=\columnwidth]{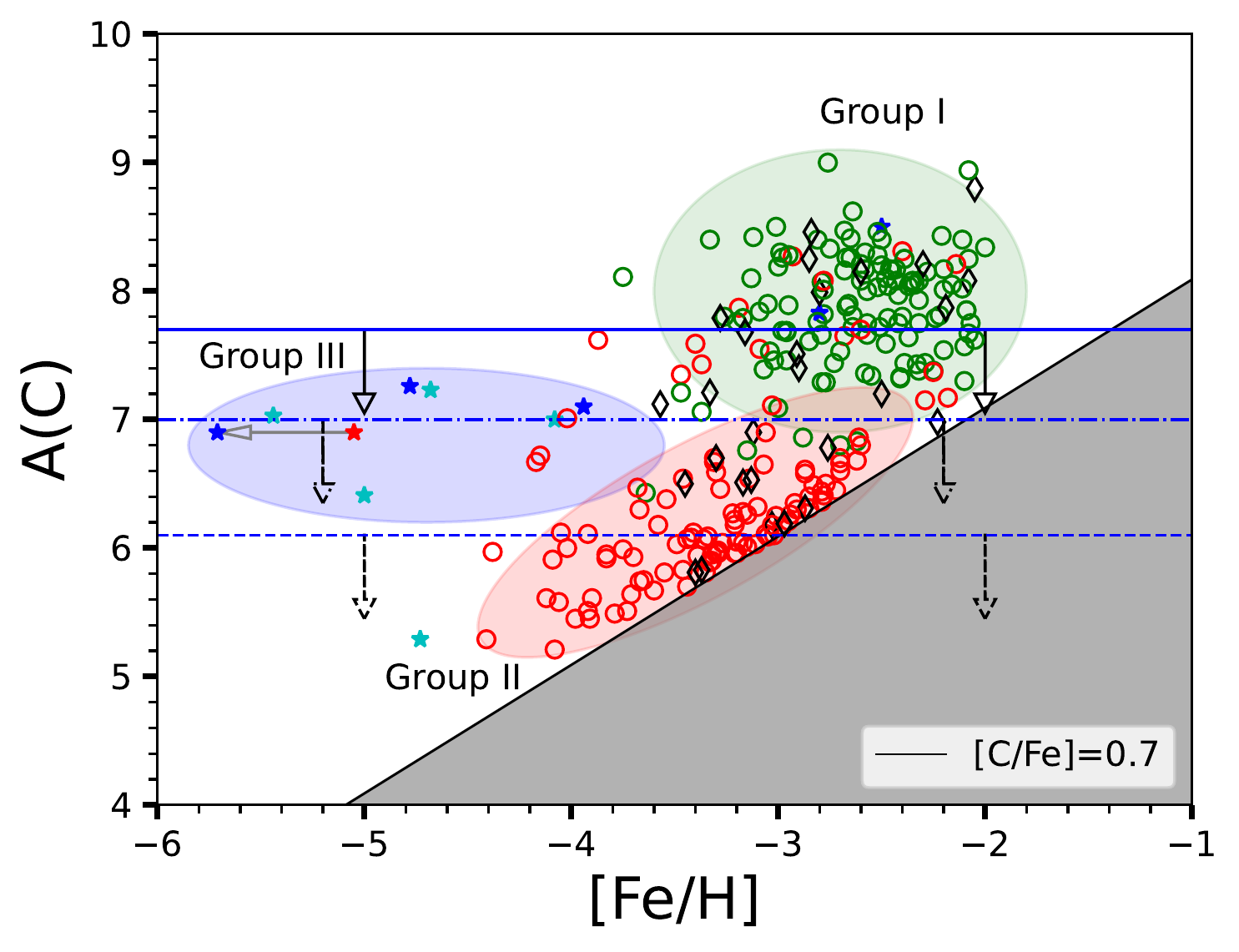}
    \caption{Same as Fig.~\ref{fig:yoon1}a but with upper limits of \A{C} from SN from non-rotating stars of $12$--$30\,\Msun$ (\textsl{blue dashed}), SN resulting from rapidly rotating QCH stars (\textsl{blue dashed-dot}), and mixing of wind ejecta from QCH stars with an external SN (\textsl{blue solid}). }   
  \label{fig:yoon_comp}
 \end{figure}
 
The low energies carried by the winds from QCH models lead to a dilution that is substantially lower than both regular and faint CCSN of energy $\gtrsim 5\times 10^{50}\,\erg$.  The values of $E_{\rm wind}\lesssim 3\times10^{49}\,\erg$ invariably leads to internal enrichment in minihaloes.  In internal enrichment, the highest enrichment (lowest effective dilution) of the collapsing star-forming cloud occurs when the SN ejecta recollapses back on the central regions of a minihalo.  This is expected to be the case for enrichment from the wind where the very low wind energies, $E_{\rm wind}$, are expected to lead to a quick recollapse of the wind ejecta back to the central regions of the minihalo to form the next generation of stars.  Lower dilution is also expected because the radius, $R_{\rm fade}$, at which a spherically symmetric blast wave fades within a fully ionized ISM, would be a factor $\gtrsim3$ smaller than that of a regular SN.  
The smaller radius causes a much faster recollapse of metals to the center of minihaloes, leading to a lower dilution and hence to a higher enrichment.  Similarly, the swept-up mass of spherically symmetric blast wave before it merges with the ISM, $M_{\rm dil}^{\rm min}$, that was found to be consistent with the lower bound of dilution in minihaloes, scales as $\sim E^{0.96}$ (Eq.~\ref{eq:magg}).  Consequently, for the wind ejecta, $M_{\rm dil}^{\rm min}$ is $\sim 400$--$1\mathord,200\,\Msun$, compared to typical values of $3.5\times 10^4\,\Msun$ for a regular SN.  On the other hand, the typical C yields in the wind ejecta of $M_{\rm C,wind}\sim 0.15$--$0.45\,\Msun$ are similar to the maximum C that can be ejected in CCSN of non-rotating Pop III stars of $\lesssim 30\,\Msun$.  As a result, the C enrichment from QCH wind ejecta can be substantially higher. Assuming a value of the mass fraction of H in the ISM, $X_{\rm H}\approx 0.75$, and the mean molecular weight of C, $\mu_{\rm C}\approx 12$, the maximum enrichment of C can reach values up to 
\begin{eqnarray}
A({\rm C}) &=& \log \left(
\frac{M_{\rm C, wind} / \mu _{\rm C}}{X_{\rm H} M_{\rm dil}^{\rm min}}
\right) + 12 \nonumber \\
&=& 7.75
+ \log \left( \frac{M_{\rm C,wind}}{0.45~\Msun} \right)
- 0.96 \log \left( \frac{E_{\rm wind}}{3\times 10^{49}~{\rm erg}} \right) \nonumber \\
&&+ 0.11 \log \left( \frac{n_0}{0.1~{\rm cm}^{-3}} \right).
\end{eqnarray}
The above expression of maximum enrichment of C for a typical $n_0=0.1~{\rm cm}^{-3}$ can also be expressed in terms of $M_{\rm C,wind}$ and $M_{\rm wind}$ as
\begin{equation}
   A({\rm C}) \approx 7.75+ \log \left( \frac{M_{\rm C,wind}/M_{\rm wind}}{0.35}   \right).
\end{equation}
Interestingly, in all the QCH models the value of $M_{\rm C,wind}/M_{\rm wind}$ is almost constant and ranges from $0.34$---$0.37$ (see Tables~\ref{tab:z25_rotation}-\ref{tab:progenitor_mass}) across all progenitor masses and initial rotation rates.
Consequently, the maximum value of $\A{C}\sim 7.75$ is also independent of progenitor mass and initial rotation rate of the QCH models and covers almost the entire range of C enrichment observed in Group III, as well as the most C-rich Group II stars in the YB diagram (see the solid blue line with down arrows in Fig.~\ref{fig:yoon_comp}).  If, on the other hand, the QCH star eventually explodes as an SN, the SN ejecta mixes with the wind ejecta resulting in high values of effective dilution discussed above.  In this case, the C produced in the core of QCH stars is ejected with a maximum yield of $\sim3\,\Msun$, which is a factor of $\sim 10$ higher than in regular SN from non-rotating stars.  This produces values of $\A{C}\lesssim 7$ that can potentially cover a considerable range of values found in Group II stars as well as some of the Group III stars (see the dashed-dot blue line with down arrows in Fig.~\ref{fig:yoon_comp}).

\begin{table*}
    \centering
     \caption{List of CEMP-no stars included in the study for comparison with theoretical models presented in section~\ref{sec:cemp-no_discussion}. Data is generated using the SAGA database \citep{SAGA}. }
    \begin{tabular}{ccccc}
    \hline\hline
       &Star&\A{C} &$\B{Fe}{H}$ & Ref. \\
       \hline
       \multirow{4}{*}{Group III} & HE1327-2326&6.21&-5.71 &\citet{frebel2008he,ezzeddine2019evidence}\\
       
        &J0815+4729&7.43&-5.49 &\citet{hernandez2020extreme}\\
        &J0023+0307 &6.30 &$<-6.6$ &\citet{aguado2018j0023+,frebel2019chemical}\\
        
        &G77-61&7.01&-4.08&\citet{plez2005analysis}\\
         \hline
        \multirow{10}{*}{Group II}&HE1338-0052&6.90&-3.06&\citet{cohen_2013}\\
        &HE1351-1049&6.7&-3.46&\citet{zhang_2011}\\
        &SDSSJ0723+3637&6.9&-3.3&\citet{aoki2012high} \\
        &CS29504-006& 6.90&-3.12&\citet{ren2012hamburg}\\ 
        &HE0055-2314& 6.66&-2.70 &\citet{cohen_2013}\\
        &HE1338-0052 &6.90&-3.0&\citet{cohen_2013} \\
        & HE0020-1741&5.78&-4.05& \citet{placco2016ApJ} \\
        &HE0015+0048&5.97&-3.08&\citet{hollek2011Apj} \\
        &HE1506-0113&6.38&-3.49&\citet{arentsen2019} \\
        &CS30314-067&5.97&-3.31&\citet{roederer2014search} \\
        \hline
    \end{tabular}
   
    \label{tab:cemp-no_stars}
\end{table*}

\begin{table*}
    \centering
     \caption{Best fit models along with the corresponding parameters $M_{\rm cut,fin}, f_{\rm cut}$, and $\alpha$ along with the dilution mass ($M{\rm_{dil,SN}}$ and $M{\rm_{dil,wind}}$) resulting for the mixing of the QCH wind ejecta with an external SN described in section~\ref{sec:scenario1}. Note that $X(Y)\equiv X\times10^Y$.}
    \begin{tabular}{ccccccccc}
    \hline\hline

        &Star&Model&$\chi^2$&$ M_{\rm cut,fin}$&$f_{\rm cut}$ & $\alpha$&$M{\rm_{dil,wind}}$&$M{\rm_{dil,SN}}$\\
        
        &&&&(\Msun)&&&($\times 10^4\,\Msun$)&($\times 10^4\,\Msun$)\\
        \hline
        
        \multirow{4}{*}{Group III}&HE1327-2326&z12&2.82&1.66 &0.06&0.98&0.72&35.5\\
       &J0815+4729&z12&3.45&1.95&8.0(-3)&0.97&0.11&3.84\\
       &J0023+0307&z15&6.22&2.24&1.0(-3)&0.61&2.24&3.50\\
      
       &G77\_61&z25&1.77&2.76&0.99&0.99&0.46&45.2\\
        
        \hline
        \multirow{4}{*}{Group II}&HE1338-0052&z17&3.47&1.78&0.52&0.93&0.33&4.32\\
        &HE1351-1049&z17&4.21&1.78&0.85&0.97&0.48&15.5\\
        &SDSSJ0723+3637&z13&0.23&1.62 &0.21&0.89&0.45&3.65\\
        &CS29504-006&z26&2.08&2.11&0.85&0.97&0.35&11.2 \\     
         \hline
    \end{tabular}
   
    \label{tab:G3_G2_ishigaki}
\end{table*}

\begin{table*}
    \centering
    \caption{The fraction $\eta_{\rm wind}$ contributed by the QCH wind ejecta for elements C, O, Ne, Na, Mg, Ca and Fe for the best-fit models listed in Table~\ref{tab:G3_G2_ishigaki}.}
    \begin{tabular}{ccccccccc}
    \hline\hline

    &Star&C&O&Ne&Na&Mg&Ca&Fe\\
    \hline
         
         \multirow{4}{*}{Group III} &HE1327-2326&0.99&0.98&0.84&0.96&0.58&0.00&0.00\\
         &J0815+4729 &1.00&1.00&1.00&1.00&0.99&0.00&0.00\\
         &J0023+0307 &0.71&0.42&0.13&0.52&0.02&0.00&0.00\\         
         &G77\_61 &0.99&0.80&0.70&0.91&0.20&0.00&0.00\\
         \hline
        \multirow{4}{*}{Group II}&HE1338-0052 &0.93&0.66&0.18&0.33&0.08&0.00&0.00\\
        &HE1351-1049&0.97&0.82&0.34&0.55&0.18&0.00&0.00\\
        &SDSSJ0723+3637&0.95&0.81&0.51&0.90&0.11&0.00&0.00\\
        &CS29504-006 &0.95&0.57&0.16&0.38&0.10&0.00&0.00\\
        \hline
    \end{tabular}
    
    \label{tab:G3_G2_frac_mixII}
\end{table*}

\begin{figure*}
    \centering
    \includegraphics[width=\textwidth]{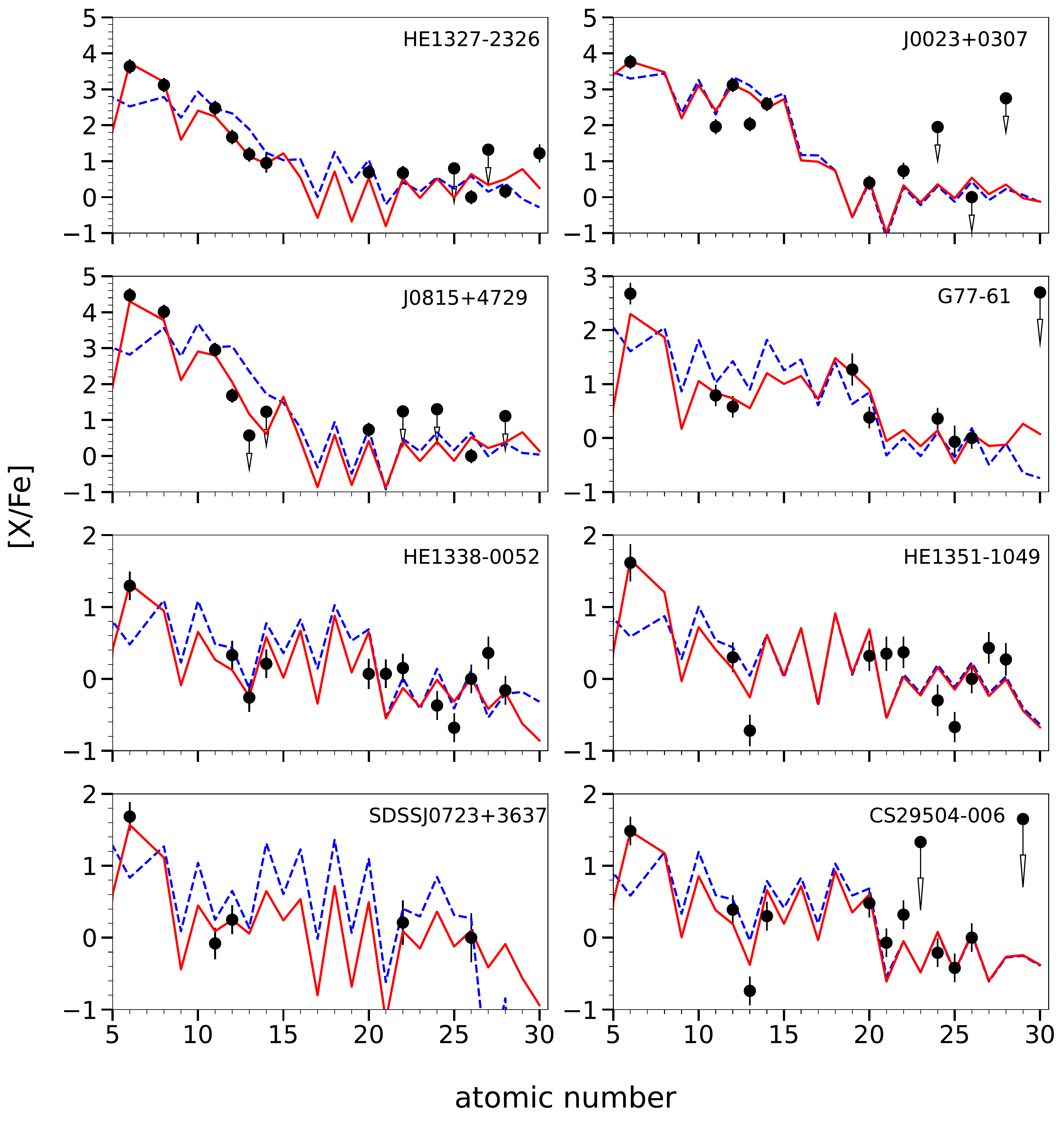}
    \caption{The elemental abundance pattern relative to Fe for the best-fit model resulting from the mixing of the QCH wind ejecta with an external SN from non-rotating progenitors (solid red) as described in section~\ref{sec:scenario1} compared to the observed abundances in CEMP-no stars from Group III and Group II (black filled circles). The best-fit models from just the non-rotating models are also plotted (blue dashed) for comparison.}
    \label{fig:GIII_GII_ishigaki}
\end{figure*}

\subsection{Formation of CEMP-no Stars} \label{sec:cemp-no_discussion}
The wind ejecta alone can only produce elements up to Mg, however, and therefore an additional contribution from an SN is required to account for heavier elements, including the Fe peak observed in CEMP-no stars.  This can be due to the mixing of the wind ejecta with either the ejecta from the SN from the same QCH star, or an SN from a different star in the same minihalo.  We consider the latter possibility first as a way to explain Group III stars with high $\A{C}\gtrsim 7$.  This situation can arise in minihaloes that host multiple massive stars and the wind ejecta can mix with gas already polluted by a previous SN. 

\subsubsection{Mixing of wind ejecta from QCH star with an external SN}\label{sec:scenario1}

First, we discuss the scenario in which wind ejecta from rapidly rotating stars is mixed with CCSN ejecta from regular non-rotating stars. Since the abundance pattern of the wind ejecta is roughly the same across all QCH models (see Section~\ref{effectofrotation_on_QCH}), we use the \texttt{z25WR$_0$60}  as our fiducial model.   We consider non-rotating models of $12$--$30\,\Msun$ with a fiducial explosion energy of $1.2\times10^{51}\,\erg$. The explosion is modelled by kinetic energy input. 
 We drive a spherically symmetric "piston" from the base of the O shell, defined as the location where the entropy per baryon first exceeds $4\,k_{\rm B}$ \citep{rauscher+2003}.  We label this mass coordinate as $M_{\rm cut, ini}$.  The CCSN ejecta for each SN progenitor star is computed using the mixing and fallback model similar to \citet{tominaga2007,ishigaki2014}.  In this prescription, following the explosive nucleosynthesis by the SN shock, all material above a mass coordinate $M_{\rm cut, fin}$ is fully ejected whereas a fraction $f_{\rm cut,}$ of the material between $M_{\rm cut, ini}$ and $M_{\rm cut, fin}$ is ejected.  Both, $M_{\rm cut, fin}$ and $f_{\rm cut}$, are treated as free parameters.  We vary $M_{\rm cut, fin }$ in steps of $0.1\,\Msun$, ranging from $M_{\rm cut, ini}$ to a maximum value that corresponds to the base of the H envelope. 
Thus, for each CCSN model, the amount of any isotope ejected by the SN depends on the values of ($M_{\rm cut,fin }$, $f_{\rm cut}$).  The mixing between the CCSN ejecta from non-rotating models and wind ejecta from the rapidly rotating star is parameterized by a single number,  
\begin{equation}
    \alpha =\frac{ M{\rm_{dil,SN}}}{M{\rm_{dil,SN}}+M{\rm_{dil,wind}}}\;,
    \label{eq:alpha}
\end{equation}
where $M{\rm_{dil,SN}}$ and $M{\rm_{dil,wind}}$ are the effective dilution mass for the CCSN and wind ejecta, respectively.  The final abundance pattern then has three parameters, i.e., $M_{\rm cut, fin}, f_{\rm cut}$, and $\alpha$.  We define the number yield, $Y_{{\rm X}_i}$, of any element,  ${\rm X}_i$, as the sum over all isotopes of the ejecta mass fractions divided by their corresponding mass numbers. 
Following the mixing of the wind and SN ejecta, the total abundance of any element relative to H can be written as 
\begin{equation}
    \frac{ N_{{\rm X}_i}}{ N_{\rm  H}}=\frac{Y_{{{\rm X}_i},{\rm wind}}}{M_{\rm H, wind}}+ \frac{Y_{{{\rm X}_i},{\rm SN}}(M_{\rm cut,fin},f_{\rm cut})}{M_{\rm H, SN}},
    \label{eq:NX_NH}
\end{equation}
where $Y_{{{\rm X}_i},{\rm wind}}$ and $Y_{{{\rm X}_i},{\rm SN}}$ are the number yield of element ${\rm X}_i$ in the wind and SN ejecta, respectively, and $M_{\rm H, wind}$ and $M_{\rm H, SN}$ are the corresponding effective mass of H that the ejecta mix with. Now, both $M_{\rm H, wind}$ and $M_{\rm H, SN}$ are simply proportional to $M{\rm_{dil,SN}}$ and $M{\rm_{dil,wind}}$, respectively, with the same proportionality constant of 0.75 corresponding to the mass fraction of H in the ISM. Combining Eq.~(\ref{eq:alpha}) and Eq.~(\ref{eq:NX_NH}), we obtain 
\begin{equation}
    \frac{ N_{{\rm X}_i}}{ N_{\rm  H}}=\frac{{ \alpha Y_{{{\rm X}_i},{\rm wind}}+\left(1-\alpha\right)\, Y_{{{\rm X}_i},{\rm SN}}(M_{\rm cut,fin},f_{\rm cut})}}{\left(1-\alpha\right)\, M_{\rm H, SN}}
    \label{eq:NX_NH_alpha}
    \;.
\end{equation}
Using Equation~(\ref{eq:NX_NH_alpha}), the abundance of any element, ${\rm X}_i$, relative to a reference element, ${\rm X_R}$, can then be written as 
\begin{equation}
    \frac{ N_{{\rm X}_i}}{ N_{\rm   X_R}}=\frac{{ \alpha Y_{{{\rm X}_i},{\rm wind}}+\left(1-\alpha\right)\, Y_{{{\rm X}_i},{\rm SN}}(M_{\rm cut,fin},f_{\rm cut})}}{{ \alpha Y_{\rm{X_R,wind}}+\left(1-\alpha\right)\, Y_{\rm X_R,SN}(M_{\rm cut,fin},f_{\rm cut})}}  
\end{equation}
where $Y_{{\rm X}_i}$ and $Y_{\rm X_R}$ are the number yield of nuclei of element ${\rm X}_i$ and ${\rm X_R}$ in the ejecta, respectively.  Following \citet{heger2010nucleosynthesis}, the best-fit model for a particular CEMP-no star is then found by minimizing the chi-square given by 
\begin{multline}
    \chi^2=\frac{1}{N+U} \left(\sum_{i=1}^N \frac{(F_i+O-D_i)^2}{\sigma_i^2} \right.\\
    \left.+\sum_{i=N+1}^{N+U} \frac{(F_i+O-D_i)^2}{\sigma_i^2}\Theta(F_i+O-D_i) \right)
\end{multline}
where $N$ is the number of elements with observed abundances, $U$ is the number of elements with upper limits, $F_i=\log\epsilon({\rm X}_i/{\rm X_R})= \log(N_{{\rm X}_i}/N_{\rm  X_R})$ is the prediction for the model for element ${\rm X}_i$, $D_i$ is the corresponding observed value in the star, $\sigma_i$ is the observed uncertainty, $\Theta(x)$ is the Heaviside function, and $O$ is the offset that minimises $\chi^2$ when all elements with upper limits are neglected given by
 \begin{equation}
     O=\left.\left(\sum_{i=1}^N\frac{(D_i-F_i)}{\sigma_i^2}\right) \middle/ \left( \sum_{i=1}^N \frac{1}{\sigma_i^2}\right)\right.
     \;.
 \end{equation}
The best-fit solution corresponds to the values of $M_{\rm cut, fin}, f_{\rm cut}$, and $\alpha$ for which $\chi^2$ is minimum.  We note that although the value of the offset, $O$, depends on the choice of reference element ${\rm X_R}$, the best-fit model and the minimum value of $\chi^2$ are independent of the choice of ${\rm X_R}$.  In our analysis, we use $\sigma_i=\max(\sigma_i,0.2)$ in order to avoid making $\chi^2$ overly sensitive to elements that have a very low value of $\sigma_i$.  In addition, we treat the total abundance of C and N as the abundance of C in order to account for the fact that the surface N in some of the low-mass stars is produced from the initial C that is converted to N via the CNO cycling which keeps the C+N abundance unchanged.   In order to calculate the dilution masses $M{\rm_{dil,SN}}$ and $M{\rm_{dil,wind}}$, the  best-fit value of the absolute abundance of any one of the elements, i.e., $\log\epsilon({\rm X}_i)_{\rm fit}$ rather than the relative abundance $\log\epsilon({\rm X}_i/{\rm X_R})_{\rm fit}$ is required.  For any element ${\rm X}_i$, $\log\epsilon({\rm X}_i)_{\rm fit}$ is related to the observed abundance $\log\epsilon({\rm X}_i)_{\rm obs}$ by 
\begin{equation}
    \log\epsilon({\rm X}_i)_{\rm fit}=\log\epsilon({\rm X}_i)_{\rm obs}+ F_{i,{\rm fit}}+O_{\rm fit}-D_i,
\end{equation}
where $F_{i,{\rm fit}}$ and $O_{\rm fit}$ are the prediction corresponding to the best-fit model and offset, respectively.
The most straightforward choice for such an element is the reference element ${\rm X_R}$ for which $F_{\rm R}=D_{\rm R}=0$ and 
\begin{equation}
    \log\epsilon({\rm X_R})_{\rm fit}=\log\epsilon({\rm X_R})_{\rm obs}+O_{\rm fit}.
    \label{eq:bestfit_loge}
\end{equation}
The corresponding dilution masses can then be calculated using $\log\epsilon({\rm X_R})_{\rm fit}$ and best-fit value of $\alpha=\alpha_{\rm fit}$ (see Appendix~\ref{appa}).
In this work, we choose Fe as the reference element.  

In order to be consistent with simulations of metal dilution by SN in minihaloes we only consider the best-fit solution for which $M{\rm_{dil,SN}}> M{\rm _{dil}^{min} }=3.5\times 10^4\,\Msun$. The value of $M{\rm _{dil}^{min}}$ corresponds to the lowest dilution found in a detailed study of metal mixing and dilution in minihaloes by \citet{chiaki2018} and the lowest among all studies as listed in \citet{magg2020minimum}. We choose eight stars with high values of \A{C} ranging from $6.2$--$7.4$ from Group II and III stars listed in Table~\ref{tab:cemp-no_stars} that cannot be explained by regular CCSN models from non-rotating stars, as mentioned earlier. 
The best-fit models and the corresponding parameters are listed in Table~\ref{tab:G3_G2_ishigaki} along with the dilution masses $M{\rm_{dil,SN}}$ and $M{\rm_{dil,wind}}$.  The values of $M{\rm_{dil,wind}}$ ranges from $\sim(1.1$--$22.4)\times10^3\,\Msun$, which is much larger than the value for the minimum dilution of $M_{\rm dil}^{\rm min}\sim 600\,\Msun$ estimated for a spherical symmetric blast wave carrying energies of $E_{\rm wind}= 1.4\times 10^{49}\,\erg$ corresponding to the total wind ejecta mass of $0.63\,\Msun$ for the \texttt{z25WR$_0$60} model (Table~\ref{tab:z25_rotation}). 
For the best-fit model for any star, for each element, the fraction of the total abundance produced by the wind ejecta $\eta_{\rm wind}$ and SN ejecta $\eta_{\rm SN}$  can be calculated (see appendix \ref{appa}) and the values for some of the key elements are listed in Table~\ref{tab:G3_G2_frac_mixII}. As can be seen clearly, C is dominantly produced by the wind in all of the stars considered here.  The wind ejecta contributes considerably to elements up to Na and Mg whereas heavier elements are exclusively produced by the SN. Figure~\ref{fig:GIII_GII_ishigaki} shows the corresponding abundance pattern compared to the best-fit model (solid-red). 
An important feature of this scenario is that because the ejecta from the non-rotating SN does not need to account for high C enrichment, which is entirely produced by the QCH wind, the values of $M{\rm_{dil,SN}}$ can be much higher than the minimum dilution mass of $3.5\E{4}\,\Msun$. This can be clearly seen from Table~\ref{tab:G3_G2_ishigaki} where four of the eight stars have $M{\rm_{dil,SN}}>\Ep{5}\,\Msun$, which is consistent with the average dilution found in detailed simulation of SN metal mixing in minihaloes \citep{chiaki2018}.

We also plot the best-fit model with just the SN ejecta, i.e., without considering the wind contribution where we again impose the criteria that $M{\rm_{dil,SN}}> M{\rm _{dil}^{min} }=3.5\times 10^4\,\Msun$.  The corresponding best-fit parameters are listed in Table~\ref{tab:non_rot} and the resulting abundance pattern is shown in Fig.\ref{fig:GIII_GII_ishigaki} as blue dashed lines.  The figures show that the SN ejecta can match the abundances of most of the elements with the clear exception of C (C+N), which is consistently underproduced.  The best-fit value of $\chi^2$ is always larger for fits with just non-rotating models compared to the models that consider mixing wind ejecta and CCSN ejecta from non-rotating stars. It is interesting to note that in Fig.\ref{fig:GIII_GII_ishigaki}, the two stars HE1327-2326 and J0815+4729 that have an observed value of O with $\B{C}{O}\sim 0.5$ which is naturally produced by the wind in contrast to best-fit models from SN ejecta alone that have $\B{C}{O}<0$. 
Although these two stars have very similar $\B{Fe}{H}$ values of $-5.71$ and $-5.49$, respectively, they have very different best-fit $M{\rm_{dil,SN}}$ values of $35.5\E4\,\Msun$ and $3.84 \E4\,\Msun$.  This is because the best-fit parameter $f_{\rm cut}$, which directly impacts the Fe yield and consequently $M{\rm_{dil,SN}}$, is 0.06 for HE1327-2326 compared to 0.008 for J0815+4729.  Interestingly, for J0815+4729, it is possible to get an equally good fit with a higher dilution mass of $12.5\E4\,\Msun$, which has a  $\chi^2=3.456$ and $f_{\rm cut}=0.02$ using the \texttt{z13} model compared to the best-fit $\chi^2=3.45$ with $f_{\rm cut}=0.008$ using model \texttt{z12}.

 \begin{table*}
    \centering
     \caption{Same as Table~\ref{tab:G3_G2_ishigaki} but for best-fit models resulting from SN from non-rotating models (Section~\ref{sec:cemp-no_discussion}).  Note that $X(Y)\equiv X\times10^Y$.}
    \begin{tabular}{ccccccc}
    \hline\hline

        &Star&Model&$\chi^2$&$ M_{\rm cut,fin}$&$f_{\rm cut}$ &$M{\rm_{dil,SN}}$\\
        
        &&&&(\Msun)&&($\times 10^4\,\Msun$)\\
        \hline
        
        \multirow{4}{*}{Group III}&HE1327-2326 & z17&8.34&2.16&8.0(-3)&10.8\\
         &J0023+0307& z16&6.01&2.32 &8.0(-4)&3.54 \\
        &J0815+4729 &z27&18.07& 2.86 & 5.0(-3)& 4.97\\
        & G77-61&z22&6.15&2.83 & 0.02&3.5\\
      
      \hline
        \multirow{9}{*}{Group II}&HE1338-0052& z26&5.65&2.79&0.48& 3.51  \\
         &HE1351-1049&z26&6.51& 2.21&0.57&10.6\\
        & CS29504-006&z26&5.13& 2.97&0.32&4.34\\
        &SDSSJ0723+3637 &z22&5.10&2.51&1.0(-5)&3.62  \\
        &HE0055-2314&z26&4.33& 2.12& 0.84&3.52 \\
        
      &HE0015+0048 &z26&2.31&2.50&0.65&5.20 \\
     &HE1506-0113 & z20&4.17&2.16&0.25&3.51\\
         & HE0020-1741 & z17&5.18&2.16&0.14&7.75\\
         &CS30314-067& z26&3.38&2.21&0.60&3.51\\
        
        \hline
       
    \end{tabular}
   
    \label{tab:non_rot}
\end{table*}

\begin{figure*}
     \centering
   \includegraphics[width=\textwidth]{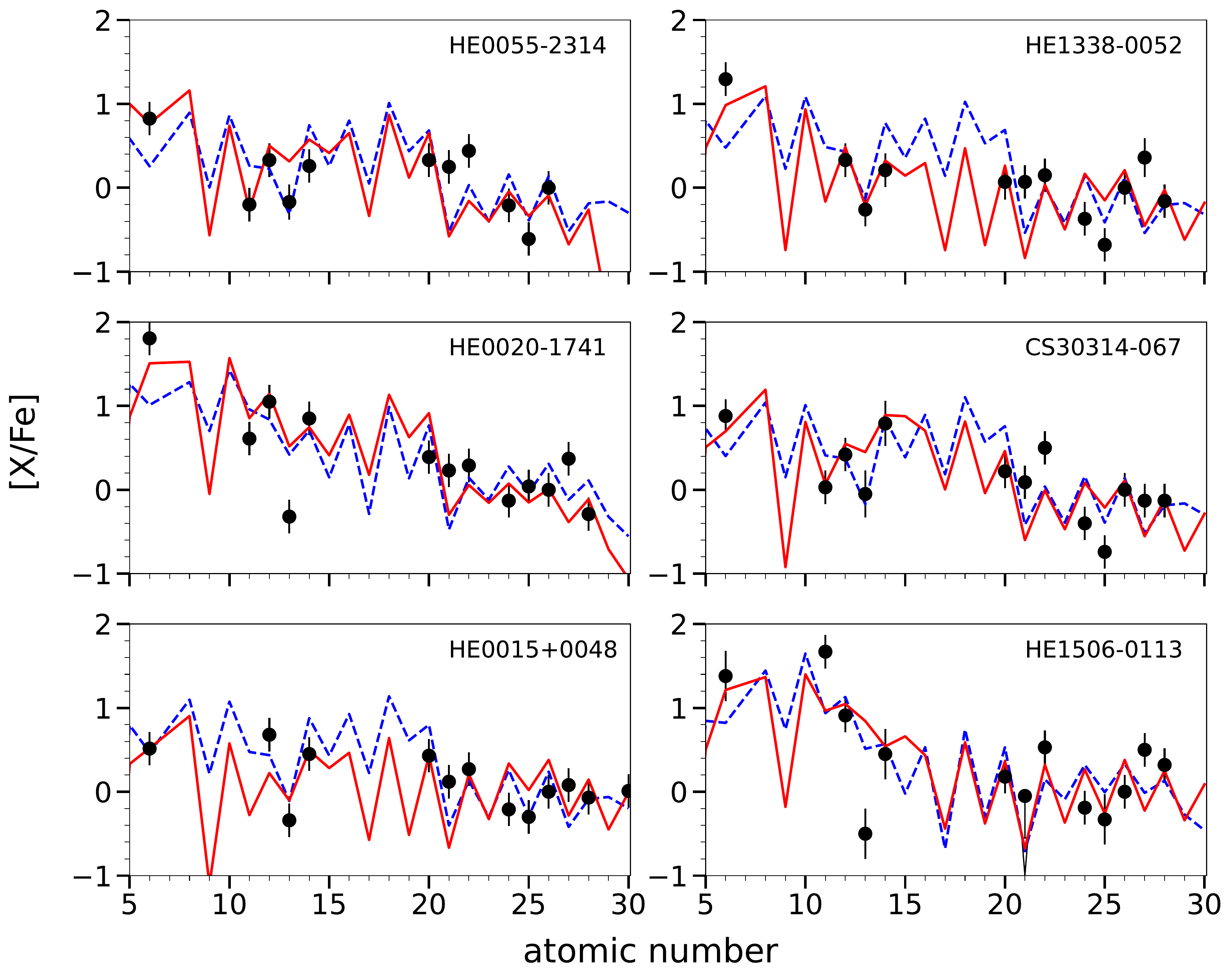}
     \caption{Same as Fig.~\ref{fig:GIII_GII_ishigaki} but for SN ejecta resulting from QCH stars of mass 20--$35\,\Msun$.}
   \label{fig:QCH_SN_wind}
 \end{figure*}

\begin{table}
    \centering
     \caption{Same as Table~\ref{tab:G3_G2_ishigaki} but for best-fit models resulting from  wind ejecta and SN ejecta from QCH star for Group II CEMP-no stars. }
    \begin{tabular}{cccccc}
    \hline\hline

        Star&Model&$\chi^2$&$ M_{\rm cut,fin}$&$f_{\rm cut}$ &$M{\rm_{dil,SN}}$\\
        
        &&&(\Msun)&&($\times 10^4\,\Msun$)\\
        \hline
        
       HE0055-2314 &  \texttt{z25WR$_0$70} &3.66& 10.29&0.31&6.97 \\
        HE1338-0052&  \texttt{z30WR$_0$60}&4.45 & 23.47&0.05& 3.93  \\
      HE0015+0048 &\texttt{z30WR$_0$60} &2.89&18.04& 0.34& 26.01 \\
       HE1506-0113 & \texttt{z30WR$_0$50}&4.78 & 28.52&0.06& 3.68\\
         HE0020-1741 & \texttt{z20WR$_0$60} &4.15&18.60 &0.02& 4.58\\
    CS30314-067& \texttt{z30WR$_0$60} &3.22&9.50 &0.13& 11.10\\
        
        \hline
       
    \end{tabular}
   
    \label{tab:QCH_explosion}
\end{table}

\subsubsection{Mixing of wind and SN ejecta from QCH star}\label{sec:scenario2}
Above we have already discussed that rapidly rotating stars can explode via a rotation-aided neutrino-driven mechanism or magnetorotational mechanism.  We explore the scenario in which the QCH star is able to undergo an explosion.  Similar to the non-rotational models, we use a fiducial explosion energy of $1.2\times 10^{51}\,\erg$ and use the mixing and fallback prescription outlined above, with its free parameters $M_{\rm cut, fin }$ and $f_{\rm cut}$.  In this case, however, as the wind ejecta is always fully mixed with the SN ejecta, the dilution masses for both are identical and thus the parameter $\alpha$ is no longer a free parameter but has a fixed value of 0.5. The rest of the procedure is identical where we minimize $\chi^2$ to find the best-fit solution.  Again, we only consider solutions for which $M{\rm_{dil,SN}}> 3.5\E{4}\,\Msun$.

We select six CEMP-no stars from Group II that cover a range of \A{C} of $5.8$--$6.9$ where $\chi^2$ for four of these models are lower, whereas two of them have similar values compared to the  CCSN models from non-rotating progenitors of $12$--$30\,\Msun$. 
The observed abundance patterns and the pattern from the best-fit models are shown in Fig.~\ref{fig:QCH_SN_wind} with the corresponding best-fit parameters listed in Table~\ref{tab:QCH_explosion}.  The best-fit models from the SN explosion of non-rotating models are also plotted for comparison.  
We find that for stars that have $\A{C}\gtrsim 6.1$, in addition to a lower or comparable $\chi^2$, the observed C abundance is fit much better with the SN ejecta from QCH star compared to non-rotating CCSN models. This is simply due to the much higher C yield in QCH stars compared to non-rotating CCSN as discussed in Section~\ref{sec:Dilution_minihalo}. In addition, because of the higher C yield, the dilution mass $M{\rm_{dil,SN}}$ for SN ejecta from QCH star is mostly higher than non-rotating CCSN while providing an equally good fit. In particular, for Group II stars with $\A{C}\lesssim6$, SN ejecta from QCH star can have $M{\rm_{dil,SN}}$ that is much higher compared to the non-rotating CCSN models as well as the minimum dilution mass of $3.5\times 10^4\,\Msun$. For example, for HE0015+0048 and CS30314-067, although the best-fit $\chi^2$ values for both QCH and normal SN models are similar, the dilution mass for SN ejecta from QCH star is $26 \times 10^{4}\,\Msun$ and $11.1\E{4}\,\Msun$, respectively, compared to $5.2 \E{4}\,\Msun$ and $3.51 \E{4}\,\Msun$ for normal non-rotating CCSN models. This is a particularly attractive feature as the effective dilution is expected to be higher than the minimum dilution with an average value of $\gtrsim \Ep{5}\,\Msun$ \citep{chiaki2018}.

\section{Summary and Outlook} \label{sec:summary}

In this paper, we simulate rapidly rotating massive metal-free stars of $20$--$35~\Msun$ with initial equatorial rotation velocities, $v_{\rm rot}$, between $40\,\%$ and $70\,\%$ of the critical speed, $v_{\rm crit}$.  Such rapidly rotating stars become QCH stars after core hydrogen depletion.  We find that in all of the QCH models, a substantial amount of CNO enriched material is ejected in the wind with C yields that are comparable to the C yields of SNe from non-rotating stars of up to $\sim0.45\,\Msun$.  The wind carries very low energy of $\lesssim 3\times10^{49}\,\erg$, much less than the typical values of $\sim 10^{51}\,\erg$ for a regular CCSN.  Consequently, the wind ejecta undergoes much lower dilution than regular SN explosions even though it has similar amounts to C. 
Importantly, we find that the ratio of the mass of C to the total mass ejected in the wind $M_{\rm C,wind}/M_{\rm wind}$ is constant across all the QCH models with a value of $\sim0.35$.
This gives rise to a high enrichment of C with a maximum value of $\A{C}\sim 7.75$ that is independent of progenitor mass and initial rotation rate.  We find that when such wind ejecta mixes with the SN ejecta from other non-rotating SN in the minihalo, the resulting abundances can easily explain the detailed abundance pattern of a wide range of CEMP-no stars belonging to Group III and Group II where the elements up to Mg are produced by the wind, whereas heavier elements up to the Fe group are produced by the SN. An important feature of this scenario is that it allows for a considerably higher dilution of the SN ejecta than the minimum dilution mass which is more consistent with the typical dilution found in detailed simulations of SN metal mixing in minihaloes \citep{chiaki2018}.

We also explored the scenario where a rapidly rotating QCH star is able to explode as SN via a magnetorotational mechanism or rotation-aided neutrino mechanism.  We find that due to the considerably larger core sizes of QCH stars that essentially cover the entire mass of the star, the maximum C yield is an order of magnitude higher than the non-rotation counterparts.  The ejecta from such explosions can result in a C enrichment of up to $\A{C}\lesssim7$.   Such an enrichment matches the detailed abundance pattern of many CEMP-no stars better than nucleosynthesis yields from CCSN of non-rotating stars along with higher dilution masses in most cases. In particular, in this scenario, a dilution much higher than the minimum dilution of $3.5\E{4}\,\Msun$ is possible compared to SN from non-rotating models for CEMP-no stars with $\A{C}\lesssim 6$. This is consistent with the average effective dilution seen in detailed studies of SN metal mixing in minihaloes \citep{chiaki2018}.

We find that rapidly rotating massive Pop III stars that reach the QCH state can explain the entire range of C enrichment in CEMP-no stars.  In particular, the omnipresence of CEMP-no stars with values of $\A{C}\gtrsim 6.1$ that cannot be explained by non-rotating models, may be an indication that a substantial fraction of the first massive stars were rapid rotators with rotation rate $\gtrsim 45\,\%$ of break-up speed. Interestingly, simulations of Pop III star formation in minihaloes by \citet{stacey2011,stacy2013} indicate that such stars could have very high rotation rates ranging from $50\,\%$ to $100\,\%$ of the break-up speed. Such rapidly rotating stars would easily reach the QCH state presented in this work.  In addition to explaining the C enrichment in CEMP-no stars, if a large fraction of Pop III stars are indeed rapid rotators, it would also increase the likelihood of rare events that are associated with rapidly-rotating stars such as hypernovae and collapsar associated with long gamma-ray bursts in the early Galaxy.  This could have an important impact on the chemical evolution of elements up to the iron peak \citep{Nomoto2006} as well as heavier elements \citep{Siegel+18,banerjee2019}.

We find that, although regular SN by themselves can explain only a fraction of CEMP-no stars with lower C enrichment of $\A{C}\lesssim 6.1$, when their ejecta in the ISM mixes with the wind ejecta from QCH stars, a wide range of observed CEMP-no star abundance patterns can be reproduced.  The wind ejecta from QCH stars can by itself enrich neighbouring ISM to high values of C and O of up to $\A{C}\lesssim 7.75$, allowing low mass stars to form directly from gas that is polluted just by the wind alone.  Such a star would likely be rare but would have a distinct abundance pattern with elements only up to $\sim$ sulphur (see Fig.~\ref{fig:wind_isotopic}).  Among the currently-known stars that come closest to such a star are  SMSS 0313-6708 \citep{Keller2014} and J0023+0307~\citep{aguado2018j0023+}.  The wind ejecta alone cannot account for the total abundance pattern of these stars, however, because Ca is detected in both.  Moreover, the extremely low value of the upper limit of Na of $\B{Na}{C}<-3.1$ in SMSS 0313-6708,  rules out the wind ejecta from QCH stars which have $\B{Na}{C}\sim -1.7$ (Fig.~\ref{fig:wind_pattern}).
Nevertheless, we expect that with a dramatic increase in the number of stars with $\B{Fe}{H}\leq -3$ from future large telescopes, some of these stars would be identified.  Future studies will also have to investigate the impact of uncertainties in the key nuclear reaction rates, such as n capture cross sections, on the $^{23}$Na yield in Pop III QCH environment.

The observational data used in Fig.~\ref{fig:yoon1}, which is directly from the original paper of \citet{yoon2016observational}, is based on 1D LTE analysis. In a recent study, \citet{Norris2019} pointed out 3D LTE/NLTE corrections to CEMP-no stars result in a reduction of \A{C} along with an increase in \B{Fe}{H}.  This results in lower values of \B{C}{Fe} and leads to a substantial decrease in the fraction of stars that are classified as CEMP-no stars that require $\B{C}{Fe}>0.7$.
Although 3D corrections highlighted by \citet{Norris2019} do affect the C abundances, the level of correction to highly C enriched stars in Group III is lower compared to Group II stars, such that even after accounting for the corrections, all of the Group III stars remain CEMP-no stars, along with some of the Group II stars.  These stars still have high values of \A{C} of up to $\sim 7.1$, which cannot be explained by regular non-rotating SN.  Moreover, the maximum value of $\A{C}\sim6.1$ that can result from a regular non-rotating SN is only reached for the most optimistic scenario in which the minimum dilution is assumed, whereas, in most simulations, the effective dilution is much larger.  For example, as mentioned earlier, in a detailed study of C enrichment from faint supernovae in minihaloes, \citet{chiaki2020seeding} find that the C enrichment for the next generations of star-forming clouds is $\A{C}\sim 4$--$5$.  Thus, even with the reduced 3D abundance of C, explaining values of $\A{C}\gtrsim 5$ in extremely metal-poor stars with $\B{Fe}{H}\lesssim -3$ is difficult to explain using only single SNe from non-rotating stars, whereas QCH stars can naturally explain the high C enrichment.

We emphasise that this study was limited to Pop III stars.  Similar rapidly rotating massive stars of very low metallicity in the early Galaxy will also produce QCH stars of similar wind and SN ejecta, although with a reduced odd-even pattern.  Such progenitors may also be relevant for explaining some of the CEMP-no stars.  Nucleosynthesis in rapidly rotating models across a range of initial metallicity along with a larger set of mass models that can be used in Galactic chemical evolution studies will be published in future works.  QCH stars from initial masses $\gtrsim40\,\Msun$ will become to pulsation-pair instability SN (PPISN; \citealt{barkat1967,hegerwoosley2002}) because almost all of the initial mass of the progenitor ultimately becomes a He core.  In addition to the wind, some of the outer regions of the star, which will have substantial amounts of C and O, will be ejected during pulsations following core C depletion.  The PPISNe ejecta have kinetic  energy ranging from $\sim 10^{48}$--$5\times10^{51}\,$ergs \citep{woosley2017}.  The ISM polluted by such ejecta could also lead to the formation of CEMP-no stars.  Due to the relatively large kinetic energies that can be comparable to regular SNe, however, the ejected mass will undergo substantial dilution and will likely lead to C enrichment of $\A{C}\lesssim 7$, similar to exploding QCH stars.  Such PPISNe ejecta could have large over-abundances in \B{C}{O} because of the partial helium burning in the He shell.  We plan to explore such models in future studies.  

Currently, there are no existing simulations of metal-mixing in minihaloes resulting from the wind and formation of the next generation of low-mass metal-poor stars.  The results of such simulations would be different from the outcomes of regular SN resulting from non-rotating stars that have been performed until now.  This is because a high C enriched gas along with higher C abundance relative to O would naturally lead to low-mass star formation due to C dust formation~\citep{chiaki2017MNRAS, chiaki2019seeding}.  Furthermore, detailed simulations of the mixing of the wind ejecta with the ejecta from other SN within the same minihalo that leads to the formation of the next generation of stars are also crucial.  We hope that our results help to motivate such studies. 

\section*{Data Availability}

Data is available upon reasonable request.

\section*{Acknowledgements}
We thank T. Sivarani and D. Karinkuzhi for their insightful discussions. P.B. was supported by the Science and Engineering Research Board Grant no SRG/2021/000673.
A.H.\ was supported by the Australian Research Council (ARC) Centre of Excellence (CoE) for Gravitational Wave Discovery (OzGrav) through project number CE170100004, by the ARC CoE for All Sky Astrophysics in 3 Dimensions (ASTRO 3D) through project number CE170100013, and by ARC LIEF grants LE200100012 and LE230100063.
\bibliography{reference}
\bibliographystyle{mnras}
\appendix

\section{Calculation of Dilution Mass from Best-fit Parameters} \label{appa}
The dilution masses, $M_{\rm dil,SN}$ and $M_{\rm dil,wind}$, can be calculated from the value of $\log \epsilon({\rm X_R)_{fit}}$ given by Eq.~(\ref{eq:bestfit_loge}) and the best-fit value $\alpha_{\rm fit}$.  Using Eq.(\ref{eq:NX_NH_alpha}) we obtain
 \begin{equation} 
     M_{\rm dil,SN}=\frac{4}{3}M_{\rm H,SN}=\frac{4}{3}\frac{N_{\rm H}}{N_{\rm  X_R}}\left(\frac{\alpha_{\rm fit}}{1-\alpha_{\rm fit}} Y_{\rm X_R,wind}+Y_{\rm X_R,SN} \right) 
     \label{eq:A1}
\end{equation}
and 
\begin{equation}
    M_{\rm  dil,wind}=\frac{1-\alpha_{\rm fit}}{\alpha_{\rm fit}} M_{\rm  dil,SN}. \label{eq:A2}
\end{equation}
The value of $N_{\rm X_R}/N_{\rm H}$ is given by 
 \begin{equation}
     \frac { N_{\rm X_R}}{ N_{\rm H}}=10^{\log\epsilon({\rm X_R})_{\rm fit}-12}
     \;. \label{eq:A3}
 \end{equation}
Substituting $N_{\rm X_R}/N_{\rm H}$ from the above equation in  Eq.~\ref{eq:A1}, we can determine  $M_{\rm dil,SN}$.  $M_{\rm dil,wind}$ can then be calculated using Eq.~\ref{eq:A2}.

\subsection{Calculation of the relative contribution of wind ejecta and SN ejecta }
We define the fraction of an element produced by the wind ejecta as
\begin{equation}
 \eta_{\rm wind}=\left. \left(\frac{Y_{{{\rm  X}_i},{\rm wind}}}{M_{\rm  dil,wind}}\right)\middle/\left( \frac{Y_{{{\rm X}_i},{\rm wind}}}{M_{\rm  dil,wind}}+\frac{Y_{{{\rm X}_i},{\rm SN}}}{M_{\rm dil,SN}}\right)\right.
 \;.
 \label{eq:etawind}
\end{equation}
The corresponding yield fraction produced by the SN is then $\eta_{\rm SN}=1-\eta_{\rm wind}$.
The above equations can be written in a compact form in terms of $\beta= Y_{{{\rm X}_i},{\rm wind}}/ Y_{{{\rm X}_i},{\rm SN}}$ and $\alpha_{\rm fit}$ as
\begin{equation}
    \eta_{\rm wind}=\frac{\alpha_{\rm fit}\beta}{1-\alpha_{\rm fit}+\alpha_{\rm fit}\beta}
\end{equation}
and
\begin{equation}
    {\rm \eta_{SN}}=\frac{1-\alpha_{\rm fit}}{1-\alpha_{\rm fit}+\alpha_{\rm fit}\beta}
    \;.
\end{equation}

\end{document}